\documentclass[lettersize,journal]{IEEEtran}
\usepackage{stmaryrd}
\usepackage{amsmath}
\usepackage{amssymb}
\usepackage{psfrag}
\usepackage{graphicx}
\usepackage{epstopdf}
\usepackage{todonotes}
\usepackage{multirow}
\usepackage{booktabs}
\usepackage{siunitx}
\usepackage{cite}
\usepackage{algorithmic}
\usepackage{algorithm}
\usepackage{svg}
\usepackage{mathtools}
\usepackage{hyperref}

\usepackage{float}

\ifCLASSOPTIONcompsoc
\usepackage[caption=false,font=normalsize,labelfont=sf,textfont=sf]{subfig}
\else
\usepackage[caption=false,font=footnotesize]{subfig}
\fi

\begin{document}
%
% paper title
% Titles are generally capitalized except for words such as a, an, and, as,
% at, but, by, for, in, nor, of, on, or, the, to and up, which are usually
% not capitalized unless they are the first or last word of the title.
% Linebreaks \\ can be used within to get better formatting as desired.
% Do not put math or special symbols in the title.
\title{RC-Struct: A Structure-based Neural Network Approach for MIMO-OFDM Detection}

\author{Jiarui Xu,
Zhou Zhou, 
Lianjun Li,
Lizhong Zheng,
and Lingjia Liu
\thanks{J. Xu, Z. Zhou, L. Li, and L. Liu are with ECE Department at Virginia Tech. 
% with the Bradley Department of Electrical and Computer Engineering, Virginia Tech, Blacksburg, VA 24061 USA.
L. Zheng is with the EECS Department at Massachusetts Institute of Technology.
% is with the Department of Electrical Engineering and Computer Science, Massachusetts Institute of Technology, Cambridge, MA 02139 USA.
% This paper was presented in part at the 2020 54th Asilomar Conference on Signals, Systems, and Computers~\cite{zhou2020learning}
The work is supported by US National Science Foundation (NSF) and Intel under grants CNS-2003059 and CNS-2002908.
This paper was presented in part at the 2021 IEEE Globecom Workshops~\cite{xu2021rc}.
}% 
% \thanks{
% % The work of J. Xu, Z. Zhou, L. Li, and L. Liu was supported by the National Science Foundation under Grant CCF1422241, Grant ECCS-1802710, Grant ECCS-1811497, and Grant CNS1811720. 
% This paper was presented in part at the 2020 54th Asilomar Conference on Signals, Systems, and Computers~\cite{zhou2020learning}.
% %\emph{(Corresponding author: Lingjia Liu.)}
% }%
%This paper is the journal version of the conference paper~\cite{zhou2020learning}.
% \thanks{Part of the work is presented in~\cite{zhou2020learning}.}
%The corresponding author is L. Liu (ljliu@ieee.org).
}
% The paper headers
% \markboth{IEEE JOURNAL ON SELECTED AREAS IN COMMUNICATIONS,~Vol.~14, No.~8, August~2015}%
% {Shell \MakeLowercase{\textit{et al.}}: Bare Demo of IEEEtran.cls for IEEE Journals}
% The only time the second header will appear is for the odd numbered pages
% after the title page when using the twoside option.
% 
% *** Note that you probably will NOT want to include the author's ***
% *** name in the headers of peer review papers.                   ***
% You can use \ifCLASSOPTIONpeerreview for conditional compilation here if
% you desire.

% If you want to put a publisher's ID mark on the page you can do it like
% this:
%\IEEEpubid{0000--0000/00\$00.00~\copyright~2015 IEEE}
% Remember, if you use this you must call \IEEEpubidadjcol in the second
% column for its text to clear the IEEEpubid mark.

% use for special paper notices
%\IEEEspecialpapernotice{(Invited Paper)}

% make the title area
\maketitle

% As a general rule, do not put math, special symbols or citations
% in the abstract or keywords.
\begin{abstract}
In this paper, we introduce a structure-based neural network architecture, namely RC-Struct, for MIMO-OFDM symbol detection.
The RC-Struct exploits the temporal structure of the MIMO-OFDM signals through reservoir computing (RC).
A binary classifier leverages the repetitive constellation structure in the system to perform multi-class detection.
The incorporation of RC allows the RC-Struct to be learned in a purely online fashion with extremely limited pilot symbols in each OFDM subframe.
The binary classifier enables the efficient utilization of the precious online training symbols and allows an easy extension to high-order modulations without a substantial increase in complexity.
Experiments show that the introduced RC-Struct outperforms both the conventional model-based symbol detection approaches and the state-of-the-art learning-based strategies in terms of bit error rate (BER).
The advantages of RC-Struct over existing methods become more significant when rank and link adaptation are adopted.
The introduced RC-Struct sheds light on combining communication domain knowledge and learning-based receive processing for 5G/5G-Advanced and Beyond.
\end{abstract}

% Note that keywords are not normally used for peerreview papers.

% For peer review papers, you can put extra information on the cover
% page as needed:
% \ifCLASSOPTIONpeerreview
% \begin{center} \bfseries EDICS Category: 3-BBND \end{center}
% \fi
%
% For peerreview papers, this IEEEtran command inserts a page break and
% creates the second title. It will be ignored for other modes.
\IEEEpeerreviewmaketitle

%%%%%%%%%%%%%%%%%%%%%%%%%%%%%%%%%%%%%%%%%%%%%%%%%%%%%%%%%%%%%%%%%%%%%%%%%%%%
%%%%%%%%%%%%%%%%%%%%%%%%%%%Section 1%%%%%%%%%%%%%%%%%%%%%%%%%%%%%%%%%%%%%%%%
%%%%%%%%%%%%%%%%%%%%%%%%%%%%%%%%%%%%%%%%%%%%%%%%%%%%%%%%%%%%%%%%%%%%%%%%%%%%

\section{Introduction}
\IEEEPARstart{M}{ultiple} input multiple output with orthogonal frequency division multiplexing (MIMO-OFDM) is the dominant waveform in modern wireless systems, such as 4G (LTE-Advanced) and 5G New Radio (NR).
The MIMO technology provides additional degrees of freedom in the spatial domain, which can be exploited through spatial multiplexing to increase the channel capacity.
To realize MIMO capacity gain, symbol detection is a crucial stage to recover multiple transmitted data streams from multiple receive antennas.

MIMO-OFDM symbol detection approaches generally fall into two categories: conventional model-based strategies and learning-based methods.
Conventional model-based symbol detection techniques usually rely on explicit modeling of the underlying system from the transmitter to the receiver. 
However, as wireless systems become more and more complicated with non-linear device components (e.g., power amplifier, low-resolution analog-to-digital converters), it is difficult to analytically model such behaviors~\cite{r2020}.
Furthermore, standard model-based approaches are usually built on top of the estimated channel state information (CSI) at the receiver.
The inaccurate modeling and CSI estimation may degrade the performance of such approaches, especially in the low signal-to-interference-plus-noise (SINR) regime. 

The recent advances of deep learning and machine learning (ML) have captured much attention on exploiting neural networks (NN) to MIMO-OFDM symbol detection~\cite{Ye2018MIMO, samuel2019learning, he2018model, khani2020adaptive}.
These learning-based approaches do not require explicit system modeling, circumventing the above issues of model-based methods.
On the other hand, learning-based strategies have their own challenges.
One major challenge is that the available online over-the-air training set is extremely limited.
For example, in 3GPP LTE/LTE-Advanced and 5G NR systems, the pilot overhead is constrained to be below 20\% to secure the spectral-efficiency~\cite{std3gpp36211}.
The limited training data may lead to the model overfitting problem for NNs and degrade the performance.

Another challenge is that NNs are difficult to be trained due to their large amount of model weights.
On one hand, a large number of model weights require more training data to learn.
On the other hand, the training process might be challenging.
For example, recurrent neural network (RNN), owing to its ability to process temporal data, has been widely explored in the symbol detection task~\cite{Lyu2020DeepNN, farsad2018neural, Liao2019DeepNN}.
However, training RNNs suffer from high computational complexity and training instability, such as the gradient vanishing and exploding problem.

One way to address these issues is to use a special type of RNN called reservoir computing (RC).
% Reservoir computing (RC), which is a special type of RNN, solves the aforementioned issues with fewer tuning weights and less training data.
The RC network consists of a RNN-based reservoir with fixed weights and a trainable output layer.
The RNN-based reservoir and light-weighted training equip RCs with the ability to process temporal sequences and can be easily trained online.
Many recent works have been devoted to investigating the effective way to utilize RC in the MIMO-OFDM symbol detection task~\cite{mosleh2017brain, zhou2019, zhou2020deep, zhou2020rcnet, li2020reservoir, zhou2020learning}.
However, such RC-based NNs are still quite generic without incorporating all available inherent structures of communication systems to realize the full potential of RC.

In this work, we introduce a new NN-based symbol detection method, RC-Struct, to leverage the inherent structural knowledge available in MIMO-OFDM systems:
%The incorporated structure knowledge are the following:
\begin{itemize}
    \item the time domain convolution and superposition due to the wireless channel;
    \item the time-frequency structure of the OFDM waveform;
    \item the repetitive structure of the modulation constellation.
\end{itemize}
% the time domain convolution and superposition operation of the wireless channel, time-frequency domain operation of OFDM systems, and the repetitive structure of the modulation constellation.
% Different from previous RC-based approaches~\cite{mosleh2017brain, zhou2019, zhou2020deep, zhou2020rcnet, li2020reservoir}, an additional frequency domain network is adopted to leverage the symmetric constellation structure after the processing of the RC.
In the time domain, we adopt RC to process sequential input as previous RC-based approaches~\cite{mosleh2017brain, zhou2019, zhou2020deep, zhou2020rcnet, li2020reservoir, zhou2020learning} allowing the underlying NN to capture the temporal dynamics embedded in the received signal and to decouple the multiple transmitted data streams for MIMO operation.
The main novelty of this work comes from the incorporation of the other two structure knowledge.
Specifically, we leverage the time-frequency structure of the OFDM waveform to make full use of the training data in these domains.
An additional NN in the frequency domain is designed to conduct symbol detection based on the repetitive structure of the modulation constellation enabling RC-Struct to conduct multi-class detection with the basic binary classifier.
%significantly improving training sample efficiency.
% In the time domain, the capability of RC to process sequential input allows the underlying NN to capture the temporal dynamics embedded in the received signal.
% In the frequency domain, the repetitive structure of the modulation constellation enables RC-Struct to conduct multi-class detection with the basic binary-class classifier to significantly improve training sample efficiency.
Due to the incorporation of the repetitive modulation structure, fewer training samples are needed for RC-Struct as opposed to generically designed NNs, significantly improving training sample efficiency.
Furthermore, RC-Struct is able to conduct symbol detection completely online in a subframe-by-subframe fashion.
In the experiments, we demonstrate that RC-Struct can outperform existing symbol detection methods for MIMO-OFDM systems in various scenarios and can conduct receive processing in a subframe-by-subframe fashion under link and rank adaptation.
The main contributions of the paper are summarized as follows:
\begin{itemize}
    % use link adapt, rank adapt, real world scenario
% \textbf{Time-frequency Structure of the OFDM Waveform}: OFDM transmit in frequency domain, waveform transmit in time domain.
% to incorporate such structure, we adopt the structu in time domain, frequency domain, time-frequency domain
\item We introduce RC-Struct for symbol detection in MIMO-OFDM systems, which takes advantage of the inherent structure knowledge of MIMO-OFDM systems.
Different from previous RC-based approaches that only utilize RC in the time domain to capture the spatiotemporal correlation of the MIMO-OFDM system, we leverage the time-frequency structure of the OFDM waveform to make full use of the available training data in different domains.
To be specific, a separate NN is introduced in the frequency domain to conduct detection/classification for transmitted symbols. 
% In MIMO-OFDM systems, the OFDM symbols transmit in the frequency domain and the waveform transmit in the time domain.
% Specifically, RC is utilized in the time domain to capture the spatio-temporal correlation of the MIMO-OFDM system, while separate NNs are utilized in the frequency domain to conduct classification for pilot symbols. 
% This architecture leverages the time-frequency structure of the OFDM waveform to make full use of the available training data.
\item The frequency domain NN takes into consideration of the repetitive structure of the modulation constellation.
%and equips it with the ability to process data online. 
%The results exhibit the advantages of using RC when compared with the conventional model-based methods.
Specifically, the multi-class classification of the high order modulation is converted into parallel binary classification tasks.
% , allowing the NN to be easily adapted to different modulation orders without re-training.
The inherent modulation constellation structure not only allows the NN to be easily adapted to different modulation orders without re-training, but also significantly improves the efficiency of utilizing the limited number of training samples.
% In this way, the inherent modulation constellation structure can help us significantly improve the training sample efficiency.
    % estimate CSI with network
    % \item The introduced RC-Struct does not assume perfect CSI. Instead, it learns the CSI with the neural network dynamically, which has revealed competitive performance than directly applying the LMMSE estimated CSI. 
    % \item The introduced RC-Struct is a subframe-based online learning approach, which can be applied with rank adaptation and link adaptation on the fly.
    % The results indicate the effectiveness and advantages of applying RC-Struct network in the MIMO-OFDM wireless systems with rank adaptation and link adaptation.
    % ad of RC based scheme in the time domain
\item RC-Struct has been designed to work in a completely online fashion under dynamic rank and link adaptation on a subframe basis for MIMO-OFDM systems. 
Extensive experiments have been conducted to demonstrate the effectiveness and advantage of our design that combines inherent structure knowledge from time, frequency, and constellation in realistic cellular environments. 
It provides a promising NN-based receive processing technique for 5G/5G-Advanced and Beyond.
%advantages of applying RC-Struct network in the MIMO-OFDM wireless 
%Experiments are conducted to show the effectiveness of adding the classifier in the frequency domain.
% ad of symmetry based learning in the frequency domain
\end{itemize}

The remainder of this paper is organized as follows:
Sec.~\ref{related_work} contains the related work and Sec.~\ref{mimo_ofdm_system} briefly discusses the structure of MIMO-OFDM systems.
Sec.~\ref{introduced_method} presents the RC-Struct while 
Sec.~\ref{evaluations} provides experimental evaluation.
%The conclusions and future research are summarized in 
Sec.~\ref{conclusion} concludes the paper.

\textbf{Notations}: In this paper, as we discuss the signal in both the time domain and frequency domain, we use lowercase and uppercase letters of $x$ and $y$ to differentiate the time domain and frequency domain signals.
Specifically, for $x$ and $y$, we use the non-bold letter for scalar and the bold letter for both vector and matrix.
For other letters, we follow the convention to use the non-bold letter for the scalar, the bold lowercase letter for the vector, and the bold uppercase letter for the matrix.
Sets get calligraphic letters.
Special sets get denoted by blackboard bold.
For example, $\mathbb{R}$ represents the real number space and $\mathbb{C}$ denotes the complex number space.

\section{Related Work}
\label{related_work}
\subsection{Deep Neural Network}
Deep neural network (DNN) methods have been recently applied to the symbol detection task in wireless systems.
State-of-the-art DNN symbol detection algorithms can be roughly divided into three categories: (1) multi-layer perception (MLP) based methods, (2) long short-term memory (LSTM) based methods, and (3) generative adversarial network (GAN) based methods, (4) convolution neural network (CNN) based methods~\cite{zhao2021deep, honkala2021deeprx}.
The first MLP-based approach, as discussed in~\cite{Ye2018MIMO}, adopted five fully-connected layers for symbol detection in OFDM systems. 
Recent advances, such as DetNet~\cite{samuel2019learning}, OAMPNet~\cite{he2018model}, MMNet~\cite{khani2020adaptive}, and HyperMIMO~\cite{goutay2020deep}, construct MLP networks by incorporating trainable parameters from conventional iterative algorithms. 
To capture the temporal information, LSTM has been utilized to learn time-memory characteristics within the data~\cite{Lyu2020DeepNN, farsad2018neural, Liao2019DeepNN}.
For GAN models, recent efforts have been devoted to treating the time-frequency channel matrix as a 2D image and estimating the channel matrix and the transmit signals jointly~\cite{Yi2020JointEst}.
More recently, DeepRx~\cite{honkala2021deeprx} is introduced to directly predict log-likelihood ratios (LLRs) of the sent bits with ResNet in the frequency domain, which has been shown to match the performance of the traditional linear minimum mean square error (LMMSE) receiver with full channel knowledge.
In~\cite{zhao2021deep}, a complex-valued network, DCCN, is adopted to directly estimate transmitted bits from the time domain signal, achieving appealing performance in different channel models.

While these DNN-based approaches achieve promising performance, they usually require a large amount of training data, making them difficult to be utilized in practice especially for 5G/5G-Advanced systems where the training is extremely limited.
In some works, such as MMNet, perfect CSI is required to train the network, which is difficult to be obtained in practice.
Different from the aforementioned DNN-based approaches, we train the network on a subframe basis, instead of training on multiple subframes, and only utilize the limited number of pilot symbols in one subframe.
Such an online learning scheme makes RC-Struct a promising approach in 5G/5G-Advanced and Beyond with dynamic transmission modes.

\subsection{Reservoir Computing in MIMO-OFDM Symbol Detection}
RC is a special type of RNN that suits temporal data processing.
Different from standard RNNs, RC can be learned with limited training data and less computational effort.
An RC consists of a reservoir part and an output layer, where the reservoir remains fixed during training and only the output layer is updated, as shown in the time domain part of Fig.~\ref{figs:rc_struct_plus_esn}.
This simple and effective training approach equips RC with the ability to efficiently learn with limited training data for symbol detection~\cite{mosleh2017brain, zhou2019, zhou2020deep, zhou2020rcnet, li2020reservoir, zhou2020learning}.
In~\cite{mosleh2017brain}, RC has been demonstrated to achieve compelling performance in online MIMO-OFDM symbol detection with limited training data.
The follow-up RC-based works show that adding a sliding window to the RC~\cite{zhou2019} and using a cascaded deep RC~\cite{zhou2020deep, zhou2020rcnet} further improve the performance.
The most recent work has focused on tracking the channel change between OFDM symbols with scattered pilots for Wi-Fi systems and the associated hardware implementation~\cite{li2020reservoir}.

Our work shares similar spirits with these efforts as we also build the NN based on RC to exploit the structural knowledge of convolution and superposition operation of the wireless channel in the time domain.
However, the key difference lies in the fact that we further leverage the time-frequency structure of OFDM and the repetitive structure of the modulation constellation in the frequency domain to construct the underlying deep NN.

%%%%%%%%%%%%%%%%%%%%%%%%%%%%%%%%%%%%%%%%%%%%%%%%%%%%%%%%%%%%%%%%%%%%%%%%%%%%
%%%%%%%%%%%%%%%%%%%%%%%%%%%Section 3%%%%%%%%%%%%%%%%%%%%%%%%%%%%%%%%%%%%%%%%
%%%%%%%%%%%%%%%%%%%%%%%%%%%%%%%%%%%%%%%%%%%%%%%%%%%%%%%%%%%%%%%%%%%%%%%%%%%%
\section{MIMO-OFDM Systems}
\label{mimo_ofdm_system}

\begin{table*}
\centering
\caption{Notations appearing in the system}
% \vspace{-2em}
% \resizebox{\linewidth}{!}{
% \begin{tabular}{ p{2.5cm} | p{5cm}}
\begin{tabular}{ l | l}
\toprule
\textbf{Symbol} & \textbf{Definition} \\
\midrule
% $T$ & Number of OFDM frames \\
$N_t$ & Number of transmitter antennas \\
$N_r$ & Number of receiver antennas \\
$N_{\mathrm{sc}}$ & Number of OFDM sub-carriers \\
$N_{c}$ & Number of channel realizations \\
$N_{\mathrm{cp}}$ & Length of Cyclic Prefix (CP) \\
$N_p$ & Number of pilot symbols in one OFDM frame (training set) \\
$N_d$ & Number of data symbols in one OFDM frame (testing set) \\ 
$N$ & Total number of symbols in one OFDM frames ($N = N_p + N_d$) \\
% $N_n$ & Number of neurons in the reservoir \\
% $L_c$ & Total number of channel delays \\
${y}_n^j(t)$ & The $t$th sample of the $n$th OFDM symbol at the $j$th receive antenna in time domain \\
${x}_n^i(t)$ & The $t$th sample of the $n$th OFDM symbol at the $i$th transmit antenna in time domain \\
$\boldsymbol{y}_n^j \in {\mathbb C}^{(N_{\mathrm{sc}} + N_{\mathrm{cp}})\times 1}$ & The $n$th OFDM symbol at the $j$th receive antenna in the time domain \\
$\boldsymbol{x}_n^i \in {\mathbb C}^{(N_{\mathrm{sc}} + N_{\mathrm{cp}})\times 1}$ & The $n$th OFDM symbol at the $i$th transmit antenna in the time domain \\
$\boldsymbol{y}_n(t) \in {\mathbb C}^{N_r\times 1}$ & The received $t$th sample of the $n$th OFDM symbol in time domain \\
$\boldsymbol{x}_n(t) \in {\mathbb C}^{N_t\times 1}$ & The transmitted $t$th sample of the $n$th OFDM symbol in time domain \\
$\boldsymbol{y}_n \in {\mathbb C}^{N_r\times (N_{\mathrm{sc}} + N_{\mathrm{cp}})}$ & The received $n$th OFDM symbol in time domain \\
$\boldsymbol{x}_n \in {\mathbb C}^{N_t\times (N_{\mathrm{sc}} + N_{\mathrm{cp}})}$ & The transmitted $n$th OFDM symbol in time domain \\
% $\boldsymbol{h}_n^{j, i} \in \mathbb{C}^{L_c}$ & The channel impulse response between $j$th receive antenna and $i$th transmit antenna \\
${Y}_n^j(k)$ & The $n$th OFDM symbol at the $j$th receive antenna and $k$th subcarrier in frequency domain \\
${X}_n^i(k)$ & The $n$th OFDM symbol at the $i$th transmit antenna and $k$th subcarrier in frequency domain \\
$\boldsymbol{Y}_n^j \in {\mathbb C}^{N_{\mathrm{sc}}\times 1}$ & The $n$th OFDM symbol at the $j$th receive antenna in frequency domain \\
$\boldsymbol{X}_n^i \in {\mathbb C}^{N_{\mathrm{sc}}\times 1}$ & The $n$th OFDM symbol at the $i$th transmit antenna in frequency domain \\
$\boldsymbol{Y}_n(k) \in {\mathbb C}^{N_r\times 1}$ & The received $n$th OFDM symbol at subcarrier $k$ in frequency domain \\
$\boldsymbol{X}_n(k) \in {\mathbb C}^{N_t\times 1}$ & The transmitted $n$th OFDM symbol at subcarrier $k$ in frequency domain \\
$\boldsymbol{Y}_n \in {\mathbb C}^{N_r\times N_{\mathrm{sc}}}$ & The received $n$th OFDM symbol in frequency domain \\
$\boldsymbol{X}_n \in {\mathbb C}^{N_t\times N_{\mathrm{sc}}}$ & The transmitted $n$th OFDM symbol in frequency domain \\
% ${N}(k) \in {\mathbb C}^{N_r\times N}$ & The additive white Gaussian noise at subcarrier $k$ \\
% ${\boldsymbol Y}(t) \in {\mathbb C}^{N_a\times N_e}$ & Received Signal \\ 
% ${{\boldsymbol {\mathcal Y}}} \in {\mathbb C}^{N_a\times N_e \times T}$  &  Stacked tensor of received signal \\  
\bottomrule
\end{tabular}
% }
\label{tab:notations}
% \vspace{-2em}
\end{table*}

\begin{figure}
\centering
\includegraphics[width=5.2cm]{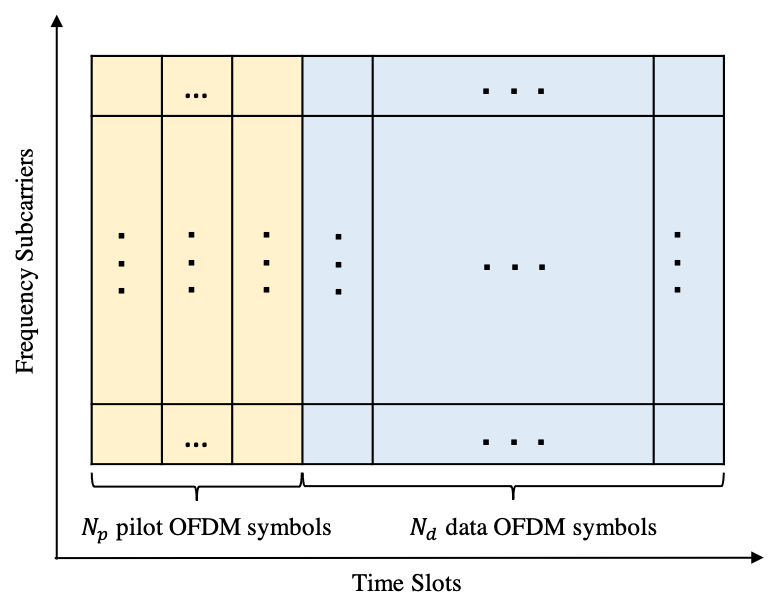}
% \vspace{-1.5em}
\caption{OFDM subframe structure and pilot patterns.}
\label{figs:ofdm_symbols}
% \vspace{-2em}
\end{figure}

In this section, we introduce the MIMO-OFDM system architecture where the notations are summarized in Tab.~\ref{tab:notations}.
In 4G/5G MIMO-OFDM systems, information is transmitted subframe by subframe where each subframe lasts for one millisecond.
For simplicity, we assume each subframe contains $N_p$ pilot symbols and $N_d$ data symbols with $N = N_p + N_d$ symbols in total, as shown in Fig.~\ref{figs:ofdm_symbols}. 
Consider a MIMO-OFDM system with $N_t$ transmit antennas, $N_r$ receive antennas, and $N_{\mathrm{sc}}$ subcarriers. 
The $n$th OFDM symbol ($n = 0, 1, \dots, N - 1$) transmitted by the $i$th transmit antenna ($i = 0, 1, \dots, N_t-1$) in the \emph{frequency domain} can be written as $\boldsymbol{X}_n^i \!\triangleq \! \left[ X_n^i(0), \!X_n^i(1), \!\dots\!,\! X_n^i(N_{\mathrm{sc}} - 1)\right]^T$, where $\boldsymbol{X}_n^i \in {\mathbb C}^{N_{\mathrm{sc}} \times 1}$ and $X_n^i(k)$ is the symbol modulated by quadrature amplitude modulation (QAM) at the $k$th subcarrier.
% \begin{align}
% \boldsymbol{X}_n^i \triangleq \left[ X_n^i(0), X_n^i(1), \dots, X_n^i(N_{\mathrm{sc}} - 1)\right]^T \in {\mathbb C}^{N_{\mathrm{sc}} \times 1}, i = 0, 1, \dots, N_t-1,
% \end{align}
% where $X_n^i(k)$ is the symbol modulated by quadrature amplitude modulation (QAM) at the $k$th subcarrier.

At the transmitter side, an inverse fast Fourier transform (IFFT) and cyclic prefix (CP) addition are applied to obtain the \emph{time domain} transmission signal $\boldsymbol{x}_n^i \!\triangleq\! \left[ x_n^i\!(0),\! x_n^i\!(1), \!\dots\!,\! x_n^i\!(\!N_{\mathrm{sc}}\! + \!N_{\mathrm{cp}} \!- \!1\!)\right]^T\!$,
% \begin{align}
% \boldsymbol{x}_n^i \triangleq \left[ x_n^i(0), x_n^i(1), \dots, x_n^i(N_{\mathrm{sc}} + N_{\mathrm{cp}} - 1)\right]^T \in {\mathbb C}^{(N_{\mathrm{sc}} + N_{\mathrm{cp}})  \times 1}, i = 0, 1, \dots, N_t-1,
% \end{align}
where $\boldsymbol{x}_n^i \in {\mathbb C}^{(N_{\mathrm{sc}} + N_{\mathrm{cp}}) \times 1}$, $N_{\mathrm{cp}}$ is the length of the CP, and $x_n^i(t)$ is the $t$th sample of the $n$th OFDM symbol at the $i$th transmit antenna in the time domain with $t = 0, 1, \dots, N_{\mathrm{sc}} + N_{\mathrm{cp}} - 1$.

The received signal at the $j$th receive antenna ($j = 0, 1, \dots, N_r$) in the \emph{time domain} can be expressed as
% \begin{align}
% \label{mimo_ofdm_signal}
% \boldsymbol{y}_n^j = \sum_{i=0}^{N_t} \boldsymbol{h_n^{j, i}} \circledast g(\boldsymbol{x_n^i}) + \boldsymbol{n}_n^j, j = 0, 1, \dots, N_r,
% \end{align}
\begin{align}
\label{mimo_ofdm_signal}
\boldsymbol{y}_n^j = \sum_{i=0}^{N_t-1} \boldsymbol{h}_n^{j, i} \circledast \phi(\boldsymbol{x}_n^i) + \boldsymbol{n}_n^j, 
\end{align}
where $\boldsymbol{y}_n^j \triangleq \left[ y_n^j(0), y_n^j(1), \dots, y_n^j(N_{\mathrm{sc}} + N_{\mathrm{cp}} - 1) \right]^T \in \mathbb{C}^{(N_{\mathrm{sc}} + N_{\mathrm{cp}}) \times 1}$, 
% \begin{align*}
%     \boldsymbol{y}_n^j \triangleq \left[ y_n^j(0), y_n^j(1), \dots, y_n^j(N_{\mathrm{sc}} + N_{\mathrm{cp}} - 1) \right]^T \in \mathbb{C}^{(N_{\mathrm{sc}} + N_{\mathrm{cp}}) \times 1}
% \end{align*}
$\boldsymbol{h}_n^{j, i} \in \mathbb{C}^{L_c}$ is the channel impulse response between $j$th receive antenna and $i$th transmit antenna with $L_c$ total number of delays; $\boldsymbol{n}_n^j$ stands for the the additive white gaussian noise (AWGN) at receiver $j$ with zero mean and noise variance $\sigma^2$; $\circledast$ represents the circular convolution operation; $\phi(\cdot)$ is the non-linear operation such as power amplifier (PA). 
% \jiarui{check the correctness here, especially for notation $n$ in the $\boldsymbol{y}_n^j$ equation.}

The corresponding received signal $\boldsymbol{Y}_n^j$ in the \emph{frequency domain} can be obtained by removing the CP and performing a fast Fourier transform (FFT), which can be denoted as $\boldsymbol{Y}_n^j \triangleq \left[ Y_n^j(0), Y_n^j(1), \dots, Y_n^j(N_{\mathrm{sc}} - 1)\right]^T \in {\mathbb C}^{N_{\mathrm{sc}} \times 1}$, 
% \begin{align}
% \boldsymbol{Y}_n^j \triangleq \left[ Y_n^j(0), Y_n^j(1), \dots, Y_n^j(N_{\mathrm{sc}} - 1)\right]^T \in {\mathbb C}^{N_{\mathrm{sc}} \times 1},
% \end{align}
where $Y_n^j(k)$ is the $n$th received symbol at the $j$th receiver and the $k$th subcarrier.

For training the RC-Struct network, we need the pilot symbols both in the time domain and frequency domain. 
For ease of discussion, we denote
\begin{gather*}
\boldsymbol{y}_n(t) \triangleq \left[ {y}_n^0(t), {y}_n^1(t), \dots, {y}_n^{N_r - 1}(t)\right]^T \in \mathbb{C}^{N_r \times 1}, \\
\boldsymbol{x}_n(t) \triangleq \left[ {x}_n^0(t), {x}_n^1(t), \dots, {x}_n^{N_t - 1}(t)\right]^T \in \mathbb{C}^{N_t \times 1}, \\
\boldsymbol{Y}_n(k) \triangleq \left[ {Y}_n^0(k), {Y}_n^1(k), \dots, {Y}_n^{N_r - 1}(k)\right]^T \in \mathbb{C}^{N_r \times 1}, \\
\boldsymbol{X}_n(k) \triangleq \left[ {X}_n^0(k), {X}_n^1(k), \dots, {X}_n^{N_t - 1}(k)\right]^T \in \mathbb{C}^{N_t \times 1},
\end{gather*}
% \begin{gather*}
% \boldsymbol{y}_{n}(t)\! \triangleq\! \left[ {y}_n^0(t), {y}_n^1(t),\! \dots, {y}_n^{N_r - 1}(t)\right]^{\!T}\! \in \!\mathbb{C}^{N_r\! \times \!1}, 
% \boldsymbol{Y}_{\!n}(k) \!\triangleq\! \left[ {Y}_n^0(k), {Y}_n^1(k), \!\dots, {Y}_n^{N_r - 1}(k)\right]^{\!T} \!\in\! \mathbb{C}^{N_r \!\times \!1}, \\
% \boldsymbol{x}_{n}(t) \!\triangleq \!\left[ {x}_n^0(t), {x}_n^1(t), \!\dots, {x}_n^{N_t - 1}(t)\right]^{\!T} \!\in\! \mathbb{C}^{N_t\! \times\! 1},
% \boldsymbol{X}_{\!n}(k) \!\triangleq \!\left[ {X}_n^0(k), {X}_n^1(k),\! \dots, {X}_n^{N_t - 1}(k)\right]^{\!T} \!\in\! \mathbb{C}^{N_t \!\times \!1},
% \end{gather*}
% \begin{gather*}
% \boldsymbol{y}_n(t) \triangleq \left[ {y}_n^0(t), {y}_n^1(t), \dots, {y}_n^{N_r - 1}(t)\right]^T, \\
% \boldsymbol{Y}_n(k) \triangleq \left[ {Y}_n^0(k), {Y}_n^1(k), \dots, {Y}_n^{N_r - 1}(k)\right]^T, \\
% \boldsymbol{x}_n(t) \triangleq \left[ {x}_n^0(t), {x}_n^1(t), \dots, {x}_n^{N_t - 1}(t)\right]^T, \\
% \boldsymbol{X}_n(k) \triangleq \left[ {X}_n^0(k), {X}_n^1(k), \dots, {X}_n^{N_t - 1}(k)\right]^T,
% \end{gather*}
where $\boldsymbol{x}_n(t) \in \mathbb{C}^{N_t \times 1}$ and $\boldsymbol{y}_n(t) \in \mathbb{C}^{N_r \times 1}$ are the transmitted and received $t$th sample of the $n$th OFDM symbol in the \emph{time domain}; 
$\boldsymbol{X}_n(k) \in \mathbb{C}^{N_t \times 1}$ and $\boldsymbol{Y}_n(k) \in \mathbb{C}^{N_r \times 1}$ are the transmitted and received $n$th OFDM symbol at the $k$th subcarrier in the \emph{frequency domain}.
Then the corresponding matrix forms are
\begin{gather*}
\boldsymbol{y}_n \triangleq \left[ \boldsymbol{y}_n(0), \boldsymbol{y}_n(1), \dots, \boldsymbol{y}_n(N_{\mathrm{sc}} \!+\! N_{\mathrm{cp}}\! -\! 1)\right] \in \mathbb{C}^{N_r \times (N_{\mathrm{sc}} \!+\! N_{\mathrm{cp}})}, \\
\boldsymbol{x}_n \triangleq \left[ \boldsymbol{x}_n(0), \boldsymbol{x}_n(1), \dots, \boldsymbol{x}_n(N_{\mathrm{sc}} \!+\! N_{\mathrm{cp}} \!-\! 1)\right] \in \mathbb{C}^{N_t \times (N_{\mathrm{sc}} \!+\! N_{\mathrm{cp}})}, \\
\boldsymbol{Y}_n \triangleq \left[ \boldsymbol{Y}_n(0), \boldsymbol{Y}_n(1), \dots, \boldsymbol{Y}_n(N_{\mathrm{sc}} - 1)\right] \in \mathbb{C}^{N_r \times N_{\mathrm{sc}}}, \\
\boldsymbol{X}_n \triangleq \left[ \boldsymbol{X}_n(0), \boldsymbol{X}_n(1), \dots, \boldsymbol{X}_n(N_{\mathrm{sc}} - 1)\right] \in \mathbb{C}^{N_t \times N_{\mathrm{sc}}},
\end{gather*}
where $\boldsymbol{x}_n \in \mathbb{C}^{N_t \times (N_{\mathrm{sc}} + N_{\mathrm{cp}})}$ and $\boldsymbol{y}_n \in \mathbb{C}^{N_r \times (N_{\mathrm{sc}} + N_{\mathrm{cp}})}$ stand for the transmitted and received $n$th OFDM symbol in the time domain, 
and $\boldsymbol{X}_n \in \mathbb{C}^{N_t \times N_{\mathrm{sc}}}$ and $\boldsymbol{Y}_n \in \mathbb{C}^{N_r \times N_{\mathrm{sc}}}$ represent the $n$th OFDM symbol in the frequency domain.

The training dataset $\{\mathcal{D}_n\}_{n = 0}^{N_p - 1}$ can be represented as
% \begin{align}
% D_n &\triangleq \left( \{\boldsymbol{y}_n(t)\}_{t=0}^{N_{\mathrm{sc}} + N_{\mathrm{cp}} - 1}, \{\boldsymbol{x}_n(t)\}_{t=0}^{N_{\mathrm{sc}} + N_{\mathrm{cp}} - 1}, \{\boldsymbol{X}_n(k)\}_{k=0}^{N_{\mathrm{sc}} - 1} \right) \\
%     &= \left( \boldsymbol{y}_n, \boldsymbol{x}_n, \boldsymbol{X}_n \right),
% \end{align}
\begin{equation}
\begin{split}
\mathcal{D}_n &\!\triangleq\! \left( \{\boldsymbol{y}_n\!(t)\}_{t=0}^{N_{\mathrm{sc}} \!+\! N_{\mathrm{cp}} \!-\! 1}\!, \{\boldsymbol{x}_n\!(t)\}_{t=0}^{N_{\mathrm{sc}} \!+ \!N_{\mathrm{cp}}\! -\! 1}\!, \{\boldsymbol{X}_n\!(k)\}_{k=0}^{N_{\mathrm{sc}} \!-\! 1} \right) \\
&\!=\! \left( \boldsymbol{y}_n, \boldsymbol{x}_n, \boldsymbol{X}_n \right),
\end{split}
\end{equation}
where $\{\boldsymbol{y}_n(t)\}_{t=0}^{N_{\mathrm{sc}} + N_{\mathrm{cp}} - 1}$ will be the input to the network; $\{\boldsymbol{x}_n(t)\}_{t=0}^{N_{\mathrm{sc}} + N_{\mathrm{cp}} - 1}$ and $\{\boldsymbol{X}_n(k)\}_{k=0}^{N_{\mathrm{sc}} - 1}$ will be the target output in the time domain and frequency domain respectively; $\boldsymbol{y}_n$, and $\boldsymbol{x}_n$, $\boldsymbol{X}_n$ are the matrix form input and target.

%%%%%%%%%%%%%%%%%%%%%%%%%%%%%%%%%%%%%%%%%%%%%%%%%%%%%%%%%%%%%%%%%%%%%%%%%%%%%%%%%%%%%%%%%%%%%%%%%%%%%%%%%%%%%%%%%%%%%%%%%%%%%%

%%%%%%%%%%%%%%%%%%%%%%%%%%%%%%%%%%%%%%%%%%%%%%%%%%%%%%%%%%%%%%%%%%%%%%%%%%%%
%%%%%%%%%%%%%%%%%%%%%%%%%%%Section 4%%%%%%%%%%%%%%%%%%%%%%%%%%%%%%%%%%%%%%%%
%%%%%%%%%%%%%%%%%%%%%%%%%%%%%%%%%%%%%%%%%%%%%%%%%%%%%%%%%%%%%%%%%%%%%%%%%%%%
\section{The Introduced Approach --- RC-Struct}
\label{introduced_method}

We introduce the RC-Struct method to exploit the properties of OFDM signals in both the time and frequency domain for symbol detection.
The introduced method is composed of two parts: RC-based time domain data-stream decoupling and equalization as well as the structure-based frequency domain NN classification.
% time domain RC and frequency domain structure-based NN.
The received signal is first decoupled and equalized by RC in the time domain and then classified by the NN in the frequency domain.
The architecture of the network is shown in Fig.~\ref{figs:rc_struct_plus_esn}.
We will first discuss the time domain RC and then focus on our structured NN in the frequency domain.
% The notations used in the time domain and frequency domain are summarized in Tab.~\ref{tab:notations_time} and Tab.~\ref{tab:notations_frequency}, respectively.

%%%%%%%%%%%%%%%%%%%%%%%%%%%%%%%%%%%%%%%%%%%%%%%%%%%%%%%%%%%%%%%%%%%%%%%%%%%%
\subsection{Time Domain: Reservoir Computing}
\subsubsection{Introduction of Reservoir Computing}

\begin{figure*}
\centering
\includegraphics[width=0.7\linewidth]{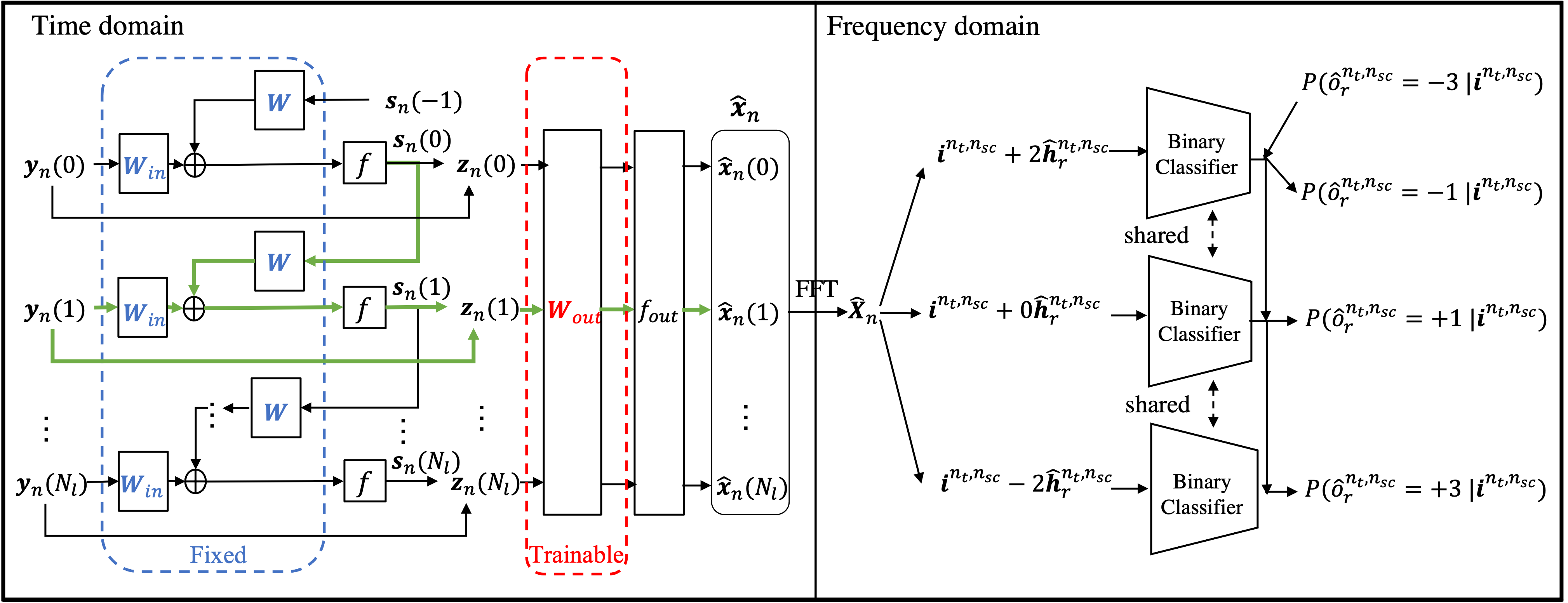}
% \vspace{-1.5em}
\caption{The architecture of RC-Struct network. 
For better visualization, we denote $N_l = N_{\mathrm{sc}}+N_{\mathrm{cp}}-1$ in the figure.
In the time domain, the fixed weights are marked as blue and the trainable weights are colored with red.
Note that all the $\boldsymbol{W}_{\mathrm{in}}$'s share the same weights and all the $\boldsymbol{W}$'s share the same weights.
The arrows for processing $\boldsymbol{y}_n(1)$ are highlighted in green to show how a specific $\boldsymbol{y}_n(t)$ is processed.
}
\label{figs:rc_struct_plus_esn}
% \vspace{-2em}
\end{figure*}

\begin{figure}
\centering
\includegraphics[width=0.6\linewidth]{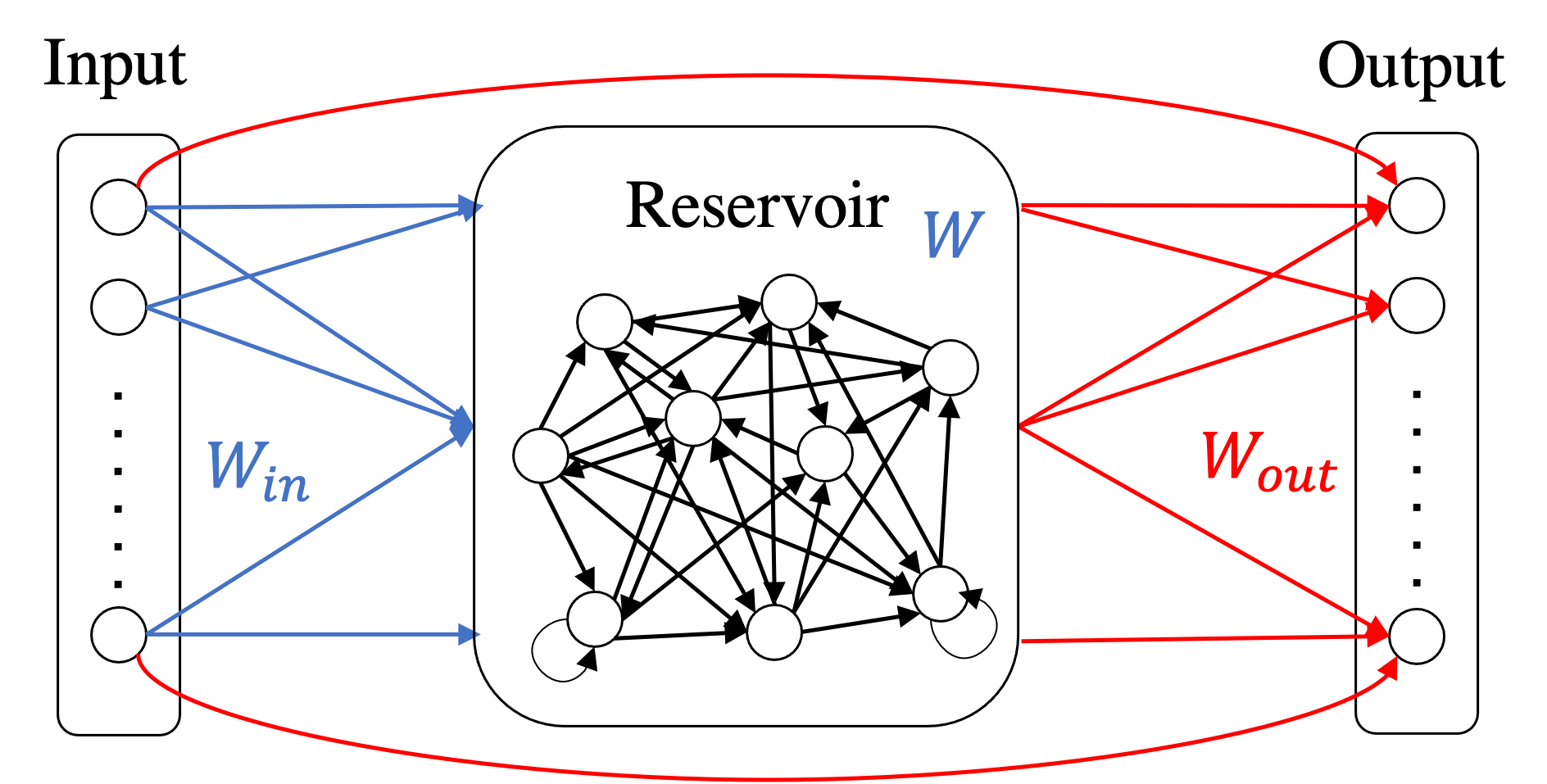}
% \vspace{-1.5em}
\caption{Echo State Network.
The fixed weights are highlighted in blue.
The trainable weights are highlighted in red.}
\label{figs:esn}
% \vspace{-2em}
\end{figure}

RNNs have been broadly applied in temporal data processing, for their special feedback and skip connections for generating history-dependent features.
However, training RNNs is inherently difficult and time-consuming.
RC is an alternative RNN-based framework that has fast-learning capability on a variety of temporal recognition tasks.
The Echo State Network (ESN)~\cite{jaeger2001echo, jaeger2004harnessing}, as a specific type of RC model, consists of the input layer, reservoir unit, and the output layer, as shown in  Fig.~\ref{figs:esn}.
Unlike conventional RNNs, the training of ESN is simple and fast.
Specifically, during training, only the weights of the output layer are learned and updated, while the weights for the input layer and the reservoir are randomly initialized and fixed.
To ensure the ESN works, the reservoir should be appropriately designed.
The RNN-based reservoir should satisfy the echo state property so that the network can asymptotically eliminate any information from the initial condition~\cite{jaeger2001echo, tanaka2019recent}.
In~\cite{lukovsevivcius2012practical}, it is shown that the echo state property can be satisfied if the spectral radius of the reservoir transition matrix is smaller than unity.
Thanks to the fast and simple training process, RC now has achieved promising performance in many tasks, such as speech recognition~\cite{triefenbach2010phoneme, verstraeten2006reservoir}, image detection~\cite{jalalvand2015real, tong2018reservoir}, wireless communication~\cite{mosleh2017brain, zhou2019, zhou2020deep, zhou2020rcnet, li2020reservoir}, etc.

% RC is a RNN-based framework that has fast-learning capability on a variety of temporal recognition tasks.
% The Echo State Network (ESN), as a specific type of RC models, employs a RNN-based reservoir with sparsely connected neurons and has been demonstrated to be effective for temporal data prediction~\cite{jaeger2001echo, jaeger2004harnessing}.
% % The Echo State Network (ESN) is a specific type of RC models, which has been demonstrated to be effective for temporal data prediction \cite{jaeger2001echo, jaeger2004harnessing}.
% % ESN employs a RNN-based reservoir with sparsely connected neurons. 
% An ESN without feedback connections has three main characteristics: the input layer, reservoir unit, and the output layer, as shown in the time domain part of Fig.~\ref{figs:rc_struct_plus_esn}.
% Unlike the weights in standard RNNs, the input matrix and the reservoir transition matrix are randomly initialized and fixed.
% Only the weights of output layer are trained in a training process.
% The reservoir is activated by a temporal sequential input and maps the activation input into a high-dimensional spatial-temporal feature space. 
% The output layer collects the features and produces outputs with a linear transformation.

\subsubsection{Reservoir Computing in RC-Struct}

RC-Struct adopts ESN in the time domain to decouple the received data streams and equalize the received signals. 
% Since ESN works in the time domain of the received signals, t
The input to the reservoir is the received signal $\boldsymbol{y}_n(t)$ and the estimated output is $\boldsymbol{\hat{x}}_n(t)$, which corresponds to the transmit signal $\boldsymbol{x}_n(t)$.
The training dataset for RC in matrix form can be represented as
\begin{align}
D_{\mathrm{rc}} & \triangleq \left( \left[ \boldsymbol{y}_0, \boldsymbol{y}_1, \dots, \boldsymbol{y}_{N_p-1} \right], 
\left[ \boldsymbol{x}_0, \boldsymbol{x}_1, \dots, \boldsymbol{x}_{N_p-1} \right]
\right).
\end{align}
For an ESN with $N_n$ neurons in the reservoir, the states and the output are expressed by the following equations:
\begin{gather}
\label{eq:state_update}
\boldsymbol{s}_n(t) = f(\boldsymbol{W}\boldsymbol{s}_n(t - 1) + \boldsymbol{W}_{\mathrm{in}} \; \boldsymbol{y}_n(t)),\\
\label{eq:output_estimate}
\boldsymbol{\hat{x}}_n(t) = f_{\mathrm{out}}(\boldsymbol{W}_{\mathrm{out}} \; \boldsymbol{z}_n(t)),
\end{gather}
where $\boldsymbol{s}_n(t) \in {\mathbb C}^{N_n \times 1} $ is the internal state estimated by the reservoir with $\boldsymbol{s}_n(-1)$ initialized as a zero vector, $\boldsymbol{z}_n(t) = \left[ \boldsymbol{s}_n(t)^T, \boldsymbol{y}_n(t)^T \right]^T \in {\mathbb C}^{(N_n + N_r) \times 1}$ is the concatenation of the internal state and the input, $\boldsymbol{W}_{\mathrm{in}} \in {\mathbb C}^{N_n \times N_r}$ is the input weight matrix, $\boldsymbol{W} \in {\mathbb C}^{N_n \times N_n}$ is the reservoir weight matrix, $\boldsymbol{W}_{\mathrm{out}} \in {\mathbb C}^{N_t \times (N_n + N_r)}$ is the output weight matrix. %, and $\boldsymbol{s}_n(-1)$ is initialized as a zero vector.
The $f(\cdot)$ and $f_{\mathrm{out}}(\cdot)$ are the activation functions for the internal units and the output units respectively.
In general, identity transformation is used for the output activation function.

In conventional RNNs, the input weights $\boldsymbol{W}_{\mathrm{in}}$, the reservoir weights $\boldsymbol{W}$, and the output weights $\boldsymbol{W}_{\mathrm{out}}$ are all learned through the backpropagation algorithm.
% The training of RNNs might suffer from the gradient exploding or vanishing problem.
Different from the conventional RNNs, in RC, only the output weight matrix $\boldsymbol{W}_{\mathrm{out}}$ is updated and all the other weights are fixed.
More importantly, the output weight matrix is learned through a closed-form least-square solution.
Specifically, RC is designed to minimize the mean square error between the network output signal $\boldsymbol{\hat{x}}_n(t)$ and the desired transmit signal $\boldsymbol{x}_n(t)$.
The regression procedure can be described as
\begin{equation}
\begin{split}
    \label{eq:rc_objective}
    \boldsymbol{\hat{W}}_{\mathrm{out}}\! &=\! \arg\!\min_{\boldsymbol{W}_{\mathrm{out}}}\! \sum_{n=0}^{N_p\!-\!1} \! \sum_{t=0}^{N_{\mathrm{cp}}\!+\!N_{\mathrm{sc}}\!-\!1}\! \| \boldsymbol{\hat{x}}_n\!(t)\! - \!\boldsymbol{x}_n\!(t) \|_{\!2}^{\!2} \\
    &=\! \arg\!\min_{\boldsymbol{W}_{\mathrm{out}}}\! \sum_{n=0}^{N_p\!-\!1} \!\| \boldsymbol{\hat{x}}_n \!- \!\boldsymbol{x}_n \|_{\!F}^{\!2} \!=\! \arg\!\min_{\boldsymbol{W}_{\mathrm{out}}}\! \| \boldsymbol{\hat{x}}\! -\! \boldsymbol{x} \|_{\!F}^{\!2}, 
\end{split}
\end{equation}
where $\boldsymbol{\hat{x}}_n = \left[ \boldsymbol{\hat{x}}_n(0), \boldsymbol{\hat{x}}_n(1), \dots, \boldsymbol{\hat{x}}_n(N_{\mathrm{sc}}+N_{\mathrm{cp}}-1) \right] \in \mathbb{C}^{N_t \times (N_{\mathrm{sc}}+N_{\mathrm{cp}})}$ is the matrix form of the output signal, $\boldsymbol{\hat{x}} = \left[ \boldsymbol{\hat{x}}_0, \boldsymbol{\hat{x}}_1, \dots, \boldsymbol{\hat{x}}_{N_p-1}\right] \in \mathbb{C}^{N_t \times N_p(N_{\mathrm{sc}}+N_{\mathrm{cp}})}$ and $\boldsymbol{x} = \left[ \boldsymbol{x}_0, \boldsymbol{x}_1, \dots, \boldsymbol{x}_{N_p-1}\right] \in \mathbb{C}^{N_t \times N_p(N_{\mathrm{sc}}+N_{\mathrm{cp}})}$ are the concatenation of the output signal matrices and target transmit signal matrices, respectively.
In the training period, the concatenated vector $\boldsymbol{z}_n(t)$ is recorded and forms the following matrix:
\begin{gather}
\boldsymbol{Z}_n \!=\! \left[ \boldsymbol{z}_n\!(0)\!,\! \boldsymbol{z}_n\!(1)\!,\! \dots\!,\! \boldsymbol{z}_n\!(N_{\mathrm{cp}}\!+\!N_{\mathrm{sc}}-1) \right] \!\in\! {\mathbb C}^{(\!N_n\! +\! N_r\!) \!\times\! (\!N_{\mathrm{cp}}\!+\!N_{\mathrm{sc}}\!)}, \\
\boldsymbol{Z} = \left[\boldsymbol{Z_0}, \boldsymbol{Z_1}, \dots, \boldsymbol{Z_{N_p-1}} \right] \in {\mathbb C}^{(N_n + N_r) \times N_p(N_{\mathrm{cp}}+N_{\mathrm{sc}})}.
\end{gather}
If $f_{\mathrm{out}}(\cdot)$ is set as the identity function, the weights are updated by the least square solution
% \begin{align}
% \hat{W}_{out} = \left[ \boldsymbol{x}_0, \boldsymbol{x}_1, \dots, \boldsymbol{x}_{N_p-1} \right]\boldsymbol{Z}^\dag,
% \end{align}
\begin{align}
\boldsymbol{\hat{W}}_{\mathrm{out}} = \boldsymbol{x}\boldsymbol{Z}^\dag,
\end{align}
where $\boldsymbol{Z}^\dag$ is the Moore-Penrose pseudo inverse of $\boldsymbol{Z}$.

At the training time, a delay parameter $d$ will be learned to compensate for the lag-effect caused by the feedback nature of RC.
The training dataset with delay $d$ can be re-written as
\begin{align}
\mathcal{D}_{\mathrm{rc}}^{(d)} \!\triangleq\! \left(\! \left[ \!\boldsymbol{y}_0\!,\! \boldsymbol{y}_1\!, \!\dots\!,\! \boldsymbol{y}_{N_p-1}\!,\! \boldsymbol{0}_{N_r \times d} \!\right], 
    \left[ \!\boldsymbol{0}_{N_t \times d},\! \boldsymbol{x}_0\!,\! \boldsymbol{x}_1\!,\! \dots\!,\! \boldsymbol{x}_{N_p\!-\!1} \!\right] 
    \!\right)\!,\!
\end{align}
where $\boldsymbol{0}$ denotes the zero matrix.
The objective function can then be written as
\begin{align}
\hat{d}, \boldsymbol{\hat{W}}_{\mathrm{out}} = \arg\min_{d} \arg\min_{\boldsymbol{W}_{\mathrm{out}}} \| \boldsymbol{\hat{x}}' - \boldsymbol{x}' \|_F^2,
\end{align}
where $\boldsymbol{x}'$ is the target transmit signal with zero inserted, and $\boldsymbol{\hat{x}}'$ stands for the RC output when reading in received signal with zero inserted.

Following the training procedure in~\cite{zhou2020rcnet}, the RC is trained with different values of $d$ in the range of $[0, N_{\mathrm{cp}}]$ with a step size of $p$. 
The optimal delay $\hat{d}$ is determined by finding the value $d$ that generates the minimal object value defined in eq.~(\ref{eq:rc_objective}).
As discussed in~\cite{zhou2019}, utilizing a sliding window when processing the input signal can further increase the short-term memory capacity of ESN.
Following this work, we adopt a sliding window with length $N_w$ to process the input data.
Specifically, the input to the ESN includes both the current $t$th sample and also the previous samples from $t-N_w+1$ to $t-1$.
% As these training tricks does not affect our introduced network in the frequency domain, we do not unfold the discussion in this paper.
More detailed discussions have been provided in our previous work~\cite{zhou2020rcnet, zhou2019}.
% The RC is trained by first finding the $\boldsymbol{\hat{W}}_{\mathrm{out}}$ and then the $\hat{d}$.
% Different $d$ values are trained and the optimal delay $\hat{d}$ is determined by minimizing the objective function with the lowest cost.
% To further increase the short term memory capacity of ESN, a sliding window with window length $N_w$ is adopted when processing the input signal~\cite{zhou2019}.

In summary, as shown in our previous work on symbol detection for MIMO-OFDM systems, RC can effectively decouple corresponding data streams and equalize the received signals in the time domain~\cite{mosleh2017brain,zhou2019,zhou2020deep,zhou2020rcnet,li2020reservoir}.
% The RC will be trained with different $d$ values and the optimal delay $\hat{d}$ is determined by minimizing the objective function (\ref{eq:rc_objective}) with the lowest cost.
%%%%%%%%%%%%%%%%%%%%%%%%%%%%%%%%%%%%%%%%%%%%%%%%%%%%%%%%%%%%%%%%%%%%%%%%%%%%
\subsection{Frequency Domain: Structure-based Neural Network}
The novel ingredient of RC-Struct comes in the frequency domain.
We introduce the structure-based NN, which leverages the time-frequency structure of the OFDM waveform as well as the repetitive structure of the modulation constellation.
The RC in the time domain and the classifier in the frequency domain are trained separately.
The output weights of RC are first learned through the closed-form least square solution.
% \begin{align}
%     \boldsymbol{\hat{x}}_n = \left[ \boldsymbol{\hat{x}}_n(0), \boldsymbol{\hat{x}}_n(1), \dots, \boldsymbol{\hat{x}}_n(N_{\mathrm{sc}} + N_{\mathrm{cp}} - 1) \right] \in \mathbb{C}^{N_t \times (N_{\mathrm{sc}} + N_{\mathrm{cp}})}
% \end{align}
After the training of RC, we obtain the RC output $\boldsymbol{\hat{x}}_n$ by processing the time domain received signal $\boldsymbol{y}_n$ with the trained RC.
The input to the frequency domain classifier is $\boldsymbol{\hat{X}}_n \in \mathbb{C}^{N_t \times N_{\mathrm{sc}}}$, which is the FFT of the RC output $\boldsymbol{\hat{x}}_n$. 
The training data of the frequency domain network is the RC output in the frequency domain $\boldsymbol{\hat{X}}_n$ and the transmitted symbols in the frequency domain $\boldsymbol{X}_n$.

% As RC has decoupled the data streams, the underlying relationship between the output of RC and the transmitted symbols in the frequency domain can be represented as
% \begin{align}
% \label{eq:effective_channel}
%     \boldsymbol{\hat{X}}_n = \boldsymbol{H}_n \odot \boldsymbol{X}_n + \boldsymbol{G}_n,
% \end{align}
% where $\boldsymbol{H}_n$ represents the effective channel for $N_t$ independent data streams after RC-based time decoupling and equalization and $\boldsymbol{G}_n$ is the additive noise after the processing of RC.
As the introduced approach works the same for all the transmitted symbols, we drop the index $n$ in the following analysis for ease of discussion.
% Then the eq.~(\ref{eq:effective_channel}) in the scalar case can be written as
Since RC has decoupled the data streams, the underlying relationship between the output of RC and the transmitted symbols in the frequency domain can be represented as
\begin{align}
    \hat{x}^{n_t,n_{\mathrm{sc}}} = h^{n_t,n_{\mathrm{sc}}} x^{n_t,n_{\mathrm{sc}}} + g^{n_t,n_{\mathrm{sc}}},
\end{align}
where $\hat{x}^{n_t,n_{\mathrm{sc}}}$ and $x^{n_t,n_{\mathrm{sc}}}$ are the $(n_t, n_{\mathrm{sc}})^{\text{th}}$ entry of the $\boldsymbol{\hat{X}}_{n}$ and $\boldsymbol{X}_{n}$, respectively; $h^{n_t,n_{\mathrm{sc}}}$ represents the effective channel coefficients after RC-based time decoupling and equalization; $g^{n_t,n_{\mathrm{sc}}}$ is the additive noise after the processing of RC.
By transforming the complex values into real values, we have
\begin{align}
    \boldsymbol{i}^{n_t,n_{\mathrm{sc}}} = \boldsymbol{\tilde{h}}_r^{n_t,n_{\mathrm{sc}}}o_r^{n_t,n_{\mathrm{sc}}} + \boldsymbol{\tilde{h}}_{\mathrm{im}}^{n_t,n_{\mathrm{sc}}}o_{\mathrm{im}}^{n_t,n_{\mathrm{sc}}} +  \boldsymbol{\tilde{g}}^{n_t,n_{\mathrm{sc}}},
\end{align}
where
\begin{gather*}
\boldsymbol{i}^{n_t,n_{\mathrm{sc}}} \!\!= \!\!
\begin{bmatrix}\!
\Re\{\!\hat{x}^{n_t,n_{\mathrm{sc}}}\!\}\\
\Im\{\!\hat{x}^{n_t,n_{\mathrm{sc}}}\!\}\!
\end{bmatrix}\!,
o_r^{n_t,n_{\mathrm{sc}}}\!\!=\!\!\Re\{\!x^{n_t,n_{\mathrm{sc}}}\!\}\!,
o_{\mathrm{im}}^{n_t,n_{\mathrm{sc}}}\!\!=\!\!\Im\{\!x^{n_t,n_{\mathrm{sc}}}\!\}\!,\!
\end{gather*}
and
\begin{gather*}
\boldsymbol{\tilde{h}}_r^{n_t,n_{\mathrm{sc}}} = 
\begin{bmatrix}
\Re\{h^{n_t,n_{\mathrm{sc}}}\}\\
\Im\{h^{n_t,n_{\mathrm{sc}}}\}
\end{bmatrix},
\boldsymbol{\tilde{h}}_{\mathrm{im}}^{n_t,n_{\mathrm{sc}}} = 
\begin{bmatrix}
-\Im\{h^{n_t,n_{\mathrm{sc}}}\}\\
\Re\{h^{n_t,n_{\mathrm{sc}}}\}
\end{bmatrix}.
\end{gather*}
% and $o_r^{n_t,n_{\mathrm{sc}}} = \Re\{x^{n_t,n_{\mathrm{sc}}}\}$, $o_{\mathrm{im}}^{n_t,n_{\mathrm{sc}}} = \Im\{x^{n_t,n_{\mathrm{sc}}}\}$.
Note that the value of $o_r^{n_t,n_{\mathrm{sc}}}$ and $o_{\mathrm{im}}^{n_t,n_{\mathrm{sc}}}$ are in the set of $\mathcal{C} = \{-2K-1, -2K+1, \dots, +2K-1, +2K+1\}$ for $M$-QAM ($M \in \{4, 16, 64, \dots\}$), where $K = \frac{\sqrt{M}-2}{2}$.
Specifically, for QPSK, the class set is $\{-1, +1\}$.
The symbol detection task can thus be formulated as a classification problem in the frequency domain.
For simplicity, we focus on discussing the real-value part.
But the same conclusion holds for the imaginary part by replacing $\hat{o}_r^{n_t,n_{\mathrm{sc}}}$ with $\hat{o}_{\mathrm{im}}^{n_t,n_{\mathrm{sc}}}$ and $\boldsymbol{\tilde{h}}_r^{n_t,n_{\mathrm{sc}}}$ with $\boldsymbol{\tilde{h}}_{\mathrm{im}}^{n_t,n_{\mathrm{sc}}}$.

\subsubsection{Network architecture}
Due to the repetitive structure of the modulation constellation, we observe that
\begin{multline}
\label{eq:repetitive_pos}
    P\{\hat{o}_r^{n_t,n_{\mathrm{sc}}} = o_r^{n_t,n_{\mathrm{sc}}}| \boldsymbol{i}^{n_t,n_{\mathrm{sc}}}\} \\
    =  P\{\hat{o}_r^{n_t,n_{\mathrm{sc}}} = +1 | \boldsymbol{i}^{n_t,n_{\mathrm{sc}}} + (-o_r^{n_t,n_{\mathrm{sc}}}+1 )\cdot\boldsymbol{\tilde{h}}_r^{n_t,n_{\mathrm{sc}}}\},
\end{multline}
\begin{multline}
    \label{eq:repetitive_neg}
    P\{\hat{o}_r^{n_t,n_{\mathrm{sc}}} = o_r^{n_t,n_{\mathrm{sc}}}| \boldsymbol{i}^{n_t,n_{\mathrm{sc}}}\} \\
    = P\{\hat{o}_r^{n_t,n_{\mathrm{sc}}} = -1 | \boldsymbol{i}^{n_t,n_{\mathrm{sc}}} + (-o_r^{n_t,n_{\mathrm{sc}}}-1 )\cdot\boldsymbol{\tilde{h}}_r^{n_t,n_{\mathrm{sc}}}\},
\end{multline}
where $\hat{o}_r^{n_t,n_{\mathrm{sc}}}$ is the estimated real-value part of the transmitted symbol.
The observation indicates that the multi-class detection can be transformed into a binary classification between $+1$ and $-1$ by shifting $\boldsymbol{i}^{n_t,n_{\mathrm{sc}}}$ with either $(-o_r^{n_t,n_{\mathrm{sc}}}+1)\cdot\boldsymbol{\tilde{h}}_r^{n_t,n_{\mathrm{sc}}}$ or $(-o_r^{n_t,n_{\mathrm{sc}}}-1)\cdot\boldsymbol{\tilde{h}}_r^{n_t,n_{\mathrm{sc}}}$.
We name the $-o_r^{n_t,n_{\mathrm{sc}}}+1$ and $-o_r^{n_t,n_{\mathrm{sc}}}-1$ as the shifting parameter.

We leverage these properties to construct the network in the frequency domain.
Due to the repetitive structure of the modulation constellation points, the multi-class classification for $M$-QAM can be divided into several binary classification processes through a shifting process.
Thus, the network only consists of a binary classifier.
The shifting process is conducted by utilizing the effective CSI between the RC processed signal and transmit signal.
Since the perfect CSI is unknown, the effective channel is estimated through LMMSE, which is denoted as $\boldsymbol{\hat{h}}_r^{n_t,n_{\mathrm{sc}}}$.
In Fig.~\ref{figs:rc_struct_plus_esn}, we show an example when the network is detecting the real-value part of the transmitted symbols modulated in $16$-QAM.
% \jiarui{add LMMSE channel estimation explaination in appendix.}
% The utilization of repetitive structure enables the network to efficiently utilize the training symbols and allows its generalization for any modulation orders.

\subsubsection{Training process}
Denote the shifting parameter as $s_r^{n_t,n_{\mathrm{sc}}}$.
At the training time, the input to the classifier is $\boldsymbol{i}^{n_t,n_{\mathrm{sc}}} + s_r^{n_t,n_{\mathrm{sc}}}\boldsymbol{\hat{h}}_r^{n_t,n_{\mathrm{sc}}}$, where $s_r^{n_t,n_{\mathrm{sc}}}$ is either $-o_r^{n_t,n_{\mathrm{sc}}}+1$ or $-o_r^{n_t,n_{\mathrm{sc}}}-1$. 
When $s_r^{n_t,n_{\mathrm{sc}}} = -o_r^{n_t,n_{\mathrm{sc}}}+1$, the binary label for the input is $b_r^{n_t,n_{\mathrm{sc}}} = +1$.
When $s_r^{n_t,n_{\mathrm{sc}}} = -o_r^{n_t,n_{\mathrm{sc}}}-1$, the binary label for the input is $b_r^{n_t,n_{\mathrm{sc}}} = -1$.
Due to this unique process of generating training labels, one pair of data $(\boldsymbol{i}^{n_t,n_{\mathrm{sc}}}, o_r^{n_t,n_{\mathrm{sc}}})$ can be used to construct two binary training samples, making the network exploit training data more efficiently.

The binary classifier consists of two linear layers and a non-linear function between the two layers.
If the estimated binary label is represented as $\hat{b}_r^{n_t,n_{\mathrm{sc}}}$, then the function approximated by the binary classifier can be written as $f_{ n_t,n_{\mathrm{sc}}}(\hat{b}_r^{n_t,n_{\mathrm{sc}}};\boldsymbol{i}^{n_t,n_{\mathrm{sc}}} + s_r^{n_t,n_{\mathrm{sc}}}\boldsymbol{\hat{h}}_r^{n_t,n_{\mathrm{sc}}})$.

\subsubsection{Testing process}
At the testing time, the input $\boldsymbol{i}^{n_t,n_{\mathrm{sc}}}$ is tested with all the possible shifting parameters in the set $\mathcal{S} = \{-2K, -2K+2, \dots, 2K-2, 2K\}$.
When transmitted with QPSK, $K=0$ and the set is $S = \{0\}$.
Following the eq.~(\ref{eq:repetitive_pos}) and eq.~(\ref{eq:repetitive_neg}), the estimated pairwise likelihood ratio for classes in set $\mathcal{C}$ can be obtained by
\begin{multline}
\label{eq:pairwise_likelihood}
\frac{{P_{n_t,n_{\mathrm{sc}}}\{\hat{o}_r^{n_t,n_{\mathrm{sc}}} = -2k+1| \boldsymbol{i}^{n_t,n_{\mathrm{sc}}}\}}}{
{P_{n_t,n_{\mathrm{sc}}}\{\hat{o}_r^{n_t,n_{\mathrm{sc}}} = -2k-1 | \boldsymbol{i}^{n_t,n_{\mathrm{sc}}}\}}} \\
= \frac{f_{n_t,n_{\mathrm{sc}}}(\hat{b}_r^{n_t,n_{\mathrm{sc}}} = +1; \boldsymbol{i}^{n_t,n_{\mathrm{sc}}} + 2k\cdot\boldsymbol{\hat{h}}_r^{n_t,n_{\mathrm{sc}}})}{
f_{n_t,n_{\mathrm{sc}}}(\hat{b}_r^{n_t,n_{\mathrm{sc}}} = -1; \boldsymbol{i}^{n_t,n_{\mathrm{sc}}} + 2k\cdot\boldsymbol{\hat{h}}_r^{n_t,n_{\mathrm{sc}}})},
\end{multline}
where $k = -K, -K+1, \dots, +K$ is the index of each shifting parameter.
For ease of discussion, we denote the likelihood ratio as
\begin{multline}
    {\mathcal L}_{+-}(\boldsymbol{i}^{n_t,n_{\mathrm{sc}}} + 2k\cdot\boldsymbol{\hat{h}}_r^{n_t,n_{\mathrm{sc}}}) \\
    := \frac{f_{n_t,n_{\mathrm{sc}}}(\hat{o}_r^{n_t,n_{\mathrm{sc}}} = +1; \boldsymbol{i}^{n_t,n_{\mathrm{sc}}} + 2k\cdot\boldsymbol{\hat{h}}_r^{n_t,n_{\mathrm{sc}}})}{
f_{n_t,n_{\mathrm{sc}}}(\hat{o}_r^{n_t,n_{\mathrm{sc}}} = -1; \boldsymbol{i}^{n_t,n_{\mathrm{sc}}} + 2k\cdot\boldsymbol{\hat{h}}_r^{n_t,n_{\mathrm{sc}}})}.
\end{multline}
Then the eq.~(\ref{eq:pairwise_likelihood}) can be expressed as
\begin{gather*}
\frac{{P_{n_t,n_{\mathrm{sc}}}\!\{\hat{o}_r^{n_t,n_{\mathrm{sc}}} \!=\! -2k\!+\!1| \boldsymbol{i}^{n_t,n_{\mathrm{sc}}}\}}}{
{P_{n_t,n_{\mathrm{sc}}}\!\{\hat{o}_r^{n_t,n_{\mathrm{sc}}} \!=\! -2k\!-\!1 | \boldsymbol{i}^{n_t,n_{\mathrm{sc}}}\}}} \!=\!
{\mathcal L}_{+-}(\boldsymbol{i}^{n_t,n_{\mathrm{sc}}} \!+\! 2k\cdot\boldsymbol{\hat{h}}_r^{n_t,n_{\mathrm{sc}}}\!).
\end{gather*}

% % \subsubsection{Multi-class classification}
% For multi-class classification, we conduct a shifting process to calculate the likelihood ratios in view of the inherent repetitive structure between the constellation points.
% The binary classifier can thus be extended to do the multi-class classification through the shifting process.
% As mentioned above, for multi-class classification with $M$-QAM, the class set is $C \in \{-2K-1, -2K+1, \dots, 2K-1, 2K+1\}$.
% The estimated pairwise likelihood ratios are expressed as
% \begin{gather*}
% \frac{P_{n_t,n_{\mathrm{sc}}}(\hat{o}_n^{n_t,n_{\mathrm{sc}}} = -2k+1| i_n^{n_t,n_{\mathrm{sc}}})}
% {P_{n_t,n_{\mathrm{sc}}}(\hat{o}_n^{n_t,n_{\mathrm{sc}}} = -2k-1 | i_n^{n_t,n_{\mathrm{sc}}})} =
% {\mathcal L}_{+-}(i_n^{n_t,n_{\mathrm{sc}}}+2ks_n^{n_t,n_{\mathrm{sc}}}),
% \end{gather*}
% where $k = -K, -K+1, \dots, +K$ is the index of each class and $s_n^{n_t,n_{\mathrm{sc}}}$ is 
% \begin{gather*}
%     s_n^{n_t,n_{\mathrm{sc}}} = \left[ \Re{\{\tilde{h}_n^{n_t,n_{\mathrm{sc}}}\}}^T, \Im{\{\tilde{h}_n^{n_t,n_{\mathrm{sc}}}\}}^T \right]^T.
% \end{gather*}
% The $\tilde{h}_n^{n_t,n_{\mathrm{sc}}}$ is the $n_t$th diagonal element of the LMMSE estimated effective channel, which is specified in Appendix. 

By collecting all the pairwise likelihood ratios, the posterior marginal estimation of each class can be written as
\begin{multline}
    P\{\hat{o}_r^{n_t,n_{\mathrm{sc}}} \!=\! -\!2k\!+\!1| \boldsymbol{i}^{n_t,n_{\mathrm{sc}}}\} \\
    =\! P\{\hat{o}_r^{n_t,n_{\mathrm{sc}}} \!=\! -\!2K\!-\!1| \boldsymbol{i}^{n_t,n_{\mathrm{sc}}}\} \!\prod_{k'=k}^{K}\!{\mathcal L}_{+-}(\boldsymbol{i}^{n_t,n_{\mathrm{sc}}} \!+\! 2k'\cdot\boldsymbol{\hat{h}}_r^{n_t,n_{\mathrm{sc}}}\!),
\end{multline}
where we assume constant probability $P\{\hat{o}_r^{n_t,n_{\mathrm{sc}}} \!=\! -\!2K\!-\!1| \boldsymbol{i}^{n_t,n_{\mathrm{sc}}}\}$.
The final class is chosen as the class that has the maximum probability.

\subsection{Complexity Analysis}
\begin{table*}[!t]
\centering
\caption{Complexity Comparison}
% \resizebox{\columnwidth}{!}{
\begin{tabular}{lcc}
\toprule
Method & Train Complexity & Test Complexity \\
\midrule
% LMMSE+LMMSE-CSI & - & $\mathcal{O}((N_{\mathrm{train}}N_{\mathrm{sc}}^2 + 7 N_{\mathrm{train}} + N_t)(N_n+N_r)N_{\mathrm{train}})$ \\
% SD+LMMSE-CSI & - & \\
RCNet & $\mathcal{O}(V(N_n + N_{\mathrm{train}} + N_t)(N_n+N_r)N_{\mathrm{train}})$ & $\mathcal{O}(V(N_n+N_r)(N_nN_{\mathrm{test}} + N_t))$  \\
RC-Struct & $\mathcal{O}(V(N_n + N_{\mathrm{train}} + N_t)(N_n+N_r)N_{\mathrm{train}} + 8N_{h}N_tN_{\mathrm{sc}}N_pN_{\mathrm{ep}})$ & $\mathcal{O}(V(N_n+N_r)(N_nN_{\mathrm{test}} + N_t)) + 4N_{h}N_tN_{\mathrm{sc}}N_d)$ \\
\bottomrule
\end{tabular}
% }
\label{tab:complex_table}
\end{table*}

In our previous work~\cite{zhou2019, shafin2018realizing, zhou2020rcnet}, we have shown that our RC-based approaches have less computation complexity than the LMMSE method when the number of subcarriers is large.
As the complexity comparison with conventional methods has been detailed discussed in our previous work, we focus on the complexity comparison between RCNet~\cite{zhou2020rcnet} and the introduced RC-Struct.
Since the costs for matrix addition and element-wise operation are negligible compared with the matrix multiplication and pseudo-inverse operation, we mainly consider the computation cost of the matrix multiplication and pseudo-inverse operation.
Note that the complexity for the multiplication of a $m\times n$ matrix and a $n \times k$ matrix is $\mathcal{O}(mnk)$, and the complexity for pseudo-inverse of a $m\times n$ matrix ($m \leq n$) is $\mathcal{O}(mn^2)$ when it is implemented by single value decomposition.

% We provide the complexity comparison of the LMMSE approach, sphere decoding (SD) approach, RCNet, and the introduced RC-Struct.
% % The LMMSE detection method is a low-complexity linear method that has been widely used in wireless communication systems.
% % SD is a maximum likelihood (ML) detection approach, which has a super high complexity and thus is seldom exploited in reality.
% % Both approaches require knowledge of the estimated CSI.
% % In this work, the LMMSE estimated channel at transmitter side is adopted for both methods.
% % RCNet~\cite{zhou2020rcnet} is a learning-based approach that leverages cascaded RC in the time domain for symbol detection.
% % In this work, we assume two RC is cascaded in RCNet.
% % In this work, we assume two RC is cascaded in RCNet, as the deeper layers lead to higher computation costs but bring marginal performance improvement.
% As the complexity of LMMSE, SD, and RCNet have already been detailed discussed in our previous work~\cite{zhou2019, shafin2018realizing, zhou2020rcnet}, we focus on the complexity analysis of the RC-Struct.

Denote the number of training samples in the time domain as $N_{\mathrm{train}} = (N_{\mathrm{cp}} + N_{\mathrm{sc}})N_p$ and the number of testing samples in the time domain as $N_{\mathrm{test}} = (N_{\mathrm{cp}} + N_{\mathrm{sc}})N_d$.
For simplicity, the complexity of the delay learning process and the sliding window process are ignored here, as they do not change the order of magnitude of the complexity.
The training complexity for RC is the sum of the state update complexity $\mathcal{O}((N_n+N_r)N_nN_{\mathrm{train}})$ and $\boldsymbol{\hat{W}}_{\mathrm{out}}$ estimation complexity $\mathcal{O}(N_{\mathrm{train}}^2(N_n+N_r) + N_tN_{\mathrm{train}}(N_n+N_r))$.
Therefore, the training complexity for RC is $\mathcal{O}((N_n + N_{\mathrm{train}} + N_t)(N_n+N_r)N_{\mathrm{train}})$.
At testing time, the forward pass includes state update process in eq.~(\ref{eq:state_update}) and the output estimation process in eq.~(\ref{eq:output_estimate}).
Thus, the testing complexity for RC is $\mathcal{O}((N_n+N_r)(N_nN_{\mathrm{test}} + N_t))$.
In RCNet, as $V$ layers of RC are cascaded, the total training and testing complexity are $\mathcal{O}(V(N_n + N_{\mathrm{train}} + N_t)(N_n+N_r)N_{\mathrm{train}})$ and $\mathcal{O}(V(N_n+N_r)(N_nN_{\mathrm{test}} + N_t))$.

As the RC-Struct builds on top of RCNet, it shares the same complexity as RCNet in the time domain.
In the frequency domain, the binary classifier with two linear layers is adopted for classification.
Suppose the number of neurons in the first layer is $N_h$.
As the input size to the first layer is $2$, the complexity for passing the first layer is $\mathcal{O}(2N_{h})$ per input sample.
As the classifier only has two classes, the number of neurons in the second layer is $2$.
Then the complexity for passing the second layer is also $\mathcal{O}(2N_{h})$ per input sample.
Therefore, the complexity for inferring the binary classifier is $\mathcal{O}(4N_{h})$ per input sample.
At training time, the number of training samples is $2N_tN_{\mathrm{sc}}N_p$, as one pair of data can construct two binary training samples.
If the number of training epochs is set as $N_{\mathrm{ep}}$, then the training complexity is $\mathcal{O}(8N_{h}N_tN_{\mathrm{sc}}N_pN_{\mathrm{ep}})$.
The number of testing samples is $N_tN_{\mathrm{sc}}N_d$.
The testing complexity is $\mathcal{O}(4N_{h}N_tN_{\mathrm{sc}}N_d)$.
The total complexity of RC-Struct is calculated by the sum of the complexity in the time domain and frequency domain.

In Tab.~\ref{tab:complex_table}, we summarized the training and testing complexities for both methods.
% We assume that RCNet has $V$ layers of RC.
% Both the LMMSE and SD exploits the LMMSE estimated CSI.
As shown in the table, RC-Struct has higher training and testing complexity than RCNet due to its extra network in the frequency domain.
However, the complexity of RC-Struct and RCNet are still in the same order of magnitude.

% Note that the complexity for the multiplication of an $m\times n$ matrix and a $n \times k$ matrix is $\mathcal{O}(mnk)$, and the complexity for pseudo-inverse of an $m\times n$ matrix ($m \leq n$) is $\mathcal{O}(mn^2)$ when implemented by single value decomposition.
% Therefore, the training complexity for RC is the sum of state update complexity $\mathcal{O}((N_n+N_r)N_nN_{\mathrm{train}})$ and output matrix estimation complexity $\mathcal{O}(N_{\mathrm{train}}^2(N_n+N_r) + N_tN_{\mathrm{train}}(N_n+N_r))$.

%%%%%%%%%%%%%%%%%%%%%%%%%%%%%%%%%%%%%%%%%%%%%%%%%%%%%%%%%%%%%%%%%%%%%%%%%%%%
%%%%%%%%%%%%%%%%%%%%%%%%%%%Section 5%%%%%%%%%%%%%%%%%%%%%%%%%%%%%%%%%%%%%%%%
%%%%%%%%%%%%%%%%%%%%%%%%%%%%%%%%%%%%%%%%%%%%%%%%%%%%%%%%%%%%%%%%%%%%%%%%%%%%

\section{Performance Evaluation}
\label{evaluations}

\subsection{Experimental Setting}
This section conducts the performance evaluation of the introduced RC-Struct for symbol detection in MIMO-OFDM systems. 
A typical MIMO-OFDM system with $N_t = 4$ transmit antennas and $N_r = 4$ receive antennas will be investigated.
Without adaptation in the transmission mode, the system will conduct a fixed rank $4$ transmission with a fixed modulation of $16$-QAM.
Gray coding is adopted for constellation mapping.
This is the typical evaluation scenario that has been conducted in the majority of NN-based symbol detection papers.
However, it is important to note that in reality, 5G/5G-Advanced systems adopt a much more dynamic transmission operation where link adaptation and rank adaptation are done on a millisecond basis~\cite{4GMIMO_OFDM}.
% if rank adaptation is not applied.
Other system parameters in the evaluation are configured as $N_{\mathrm{sc}} = 1024, N_{\mathrm{cp}} = 160, N_p = 4, N_d = 16, N = 20$.
The training overhead of the underlying system is $20\%$ as opposed to other learning-based methods that require a prohibitively large training set.
Meanwhile, it also complies with the pilot occupancy requirement specified in the 3GPP LTE/LTE-Advanced and 5G NR systems~\cite{std3gpp36211}.
% The training overhead $20\%$ satisfies the pilot occupancy requirement in the 3GPP LTE / LTE-Advanced and 5G systems~\cite{std3gpp36211}.
For simplicity, the first $N_p$ OFDM symbols of each subframe are set to be pilot symbols, while the rest $N_d$ OFDM symbols are set to be data symbols, as shown in Fig.~\ref{figs:ofdm_symbols}.
The pilot symbols are randomly chosen from the QAM constellation.
It is important to note that similar to our previous work~\cite{zhou2019}, RC-Struct can be readily extended to scattered pilot patterns where the training overhead can be significantly reduced.
The channel coefficients are generated using the QuaDRiGa channel simulator~\cite{jaeckel2014quadriga} following the 3GPP 3D MIMO model defined in~\cite{study3d3gpp}.
In this paper, we focus on the scenario when users are in low mobilities.
To be specific, the user speed for generating the channel is set as $5$ km/h.
Our work on high mobility users can be found in~\cite{zhou2021otfs}.

The RC used in the experiments has $N_n=16$ neurons and the input window length is $N_w = 128$.
The step size for searching the optimal delay parameter is set as $p=5$.
Following RCNet, two cascaded RC are adopted in the time domain.
% we adopt an input window length $N_w = 64$ for RC to increase the short term memory capacity of ESN, as discussed in~\cite{zhou2019}.
% The state transition matrix $\boldsymbol{W}$ is randomly generated and has been set to have spectral radius smaller than 1 to satisfy the echo state property~\cite{jaeger2001echo}.
% The input weight matrix $W_{in}$ has weights generated from a uniform distribution.
% The optimal delay parameter is searched between the 0 and 140 with a step size 5.
In the frequency domain, the binary classifier consists of two linear layers, each of which has $128$ neurons.
The two layers are connected with hyperbolic tangent non-linear function.
% a multilayer perceptron (MLP) with two linear layers is built to do the binary classification, where the two layers are connected with hyperbolic tangent non-linear function and each of the layers has $64$ neurons.
The weights of the hidden layers are initialized with Xavier weight initialization~\cite{glorot2010understanding}.
The network is updated with SGD methods with the learning rate of $0.01$ and the momentum of $0.001$.
The underlying network is trained for $800$ epochs.
During training, seven resource block groups (RBGs), each of which consists of $12$ subcarriers, are combined together to train a single classification network in order to reduce the computation complexity.

In our experiments, we compare our RC-Struct with state-of-the-art learning algorithms: RCNet and MMNet.
For conventional model-based signal processing strategies, we focus on the popular LMMSE-based and sphere decoding-based algorithms.
Overall, the compared schemes include: 
(1) \emph{LMMSE+LMMSE-CSI}: The linear decoder that exploits the LMMSE-based symbol detector with the LMMSE-based estimated CSI;
(2) \emph{SD+LMMSE-CSI}: The sphere decoding-based method for symbol detection using the LMMSE-based estimated CSI~\cite{ghasemmehdi2011faster};
(3) \emph{MMNet}: The MMNet network designed for symbol detection under arbitrary channel matrices using the LMMSE-based estimated CSI~\cite{khani2020adaptive}.
To achieve the best performance, we adopt the offline setting in~\cite{khani2020adaptive} and train the network for each subcarrier with $500$ iterations.
Note that the network for each subcarrier has trained for $1000$ iterations in~\cite{khani2020adaptive}.
However, as we have $1024$ subcarriers, it is time-consuming to train the network for each subcarrier with $1000$ iterations.
Moreover, we observe that the training BER does not change anymore after $500$ iterations and thus choose to train for $500$ iterations to avoid possible overfitting;
% The online scheme refers to the training strategy described in the paper, where the NN for the first subcarrier is trained from scratch for $1000$ iterations and the NNs for the remaining subcarriers are fine-tuned on their corresponding previous NNs with $3$ additional iterations;
(4) \emph{RCNet}: The RC-based approach with $2$ cascaded RC in time domain~\cite{zhou2020rcnet}. All the parameters are set the same as the time domain RC in RC-Struct;
(5) \emph{RC-Struct}: The introduced method with LMMSE-based estimated shifting parameters.
% Overall, the compared schemes are summarized in the following: 
% \begin{itemize}
% \item \emph{LMMSE+LMMSE-CSI}: The linear decoder that exploits the LMMSE-based symbol detector with the LMMSE-based estimated CSI.
% \item \emph{SD+LMMSE-CSI}: The sphere decoding-based method for symbol detection using the LMMSE-based estimated CSI~\cite{ghasemmehdi2011faster}.
% \item \emph{MMNet}: The MMNet network designed for symbol detection under arbitrary channel matrices using the LMMSE-based estimated CSI~\cite{khani2020adaptive}.
% To achieve the best performance, we adopt the offline setting in~\cite{khani2020adaptive} and train the network for each subcarrier with $500$ iterations.
% Note that the network for each subcarrier has trained for $1000$ iterations in~\cite{khani2020adaptive}.
% However, as we have $1024$ subcarriers, it is time-consuming to train the network for each subcarrier with $1000$ iterations.
% Moreover, we observe that the training BER does not change anymore after $500$ iterations and thus choose to train for $500$ iterations to avoid possible overfitting.
% \item \emph{RCNet}: The RC-based approach with $2$ cascaded RC in time domain~\cite{zhou2020rcnet}. All the parameters are set the same as the time domain RC in RC-Struct.
% \item \emph{RC-Struct}: The introduced method with LMMSE-based estimated shifting parameters.
% \end{itemize}
We notice that initializing MMNet with pre-trained offline weights can lead to better performance than training from scratch.
However, when dynamic link adaptation and/or rank adaptation are adopted, it is almost impossible to initialize the network with pre-trained weights due to the underlying model mismatch.
Since this paper focuses on realistic 5G/5G-Advanced transmission operations with dynamic rank and link adaptation, we do not incorporate the results for MMNet with pre-trained weights.
% \subsection{Implementation details}
% The RC in the experiments has $N_n=16$ neurons using the windowed-ESN structure~\cite{zhou2019} with an input window length $64$.
% % The optimal delay parameter is searched between the 0 and 140 with a step size 5.
% In the frequency domain, a multilayer perceptron (MLP) with two linear layers is built to do the binary classification, where the two layers are connected with tanh non-linear function and each of them has $64$ neurons.
% An extra linear layer is used to keep track of the CSI.

% We initialize the MLP classifier with the offline weights that are trained with randomly generated samples from $Y=X+N$, which makes the network a binary nearest neighbor classifier.
% It is noteworthy that no real data is employed in the offline training and the offline weights can be generated with free cost.
% The linear layer for estimating the channel is initialized with the linear minimum mean square error (LMMSE) estimated channel.
% During training, 7 RBGs are grouped together to train a single classification network in order to reduce the computation complexity.
% We perform the $100$-epoch network training with a learning rate of $0.1$ and a momentum of $0.01$.
% The weights of the classifier are fixed and only the channel estimation layer is trained in the online training stage.

\subsection{Key Performance Indicators (KPI) / Performance Metrics}

In this paper, we use BER as the key performance indicator (KPI) / evaluation metric when no adaptation is applied.
When link and rank adaptations are adopted, BER along is not suitable as the KPI since it neglects the order of different modulation as well as the rank of the transmission. 
Accordingly, we adopt the RawBER as the KPI / metric to evaluate the underlying system performance~\cite{peng2007adaptive} so that data streams with different modulation orders will be factored in the performance evaluation. 
% and channels with different ranks
% To better evaluate the performance of the symbol detection with link and rank adaptation, we adopt the RawBER to evaluate the performance~\cite{peng2007adaptive} so that data streams with different modulation order can contribute equally to the resulting performance. 
To be specific, the RawBER is defined as
% \begin{align}
% RawBER = \frac{\sum_{j=1}^{N_{\mathrm{ds}}}b_jBER_j}{N_{ds}\sum_{j=1}^{N_{ds}}b_j},
% \end{align}
\begin{align}
RawBER = \frac{\sum_{j=1}^{N_{\mathrm{ds}}}b_jBER_j}{\sum_{j=1}^{N_{\mathrm{ds}}}b_j},
\end{align}
where $N_{\mathrm{ds}}$ is the number of data streams, $b_j$ is the number of bits per symbol and $BER_j$ is the BER of the detector on the $j$th data stream.
Note that RawBER is equivalent to BER if all streams are modulated with the same modulation order.
% Note that RawBER takes care of both the link and the rank of the underlying transmission.
% Furthermore, it is equivalent to BER if all streams are modulated with the same modulation order.
% The $N_{ds}$ is divided here to penalize channels with different ranks.
% We will use BER as the evaluation metric when no adaptation is applied and utilize RawBER when applying link or rank adaptation.
% \jiarui{Not sure if we need to explain this. "The $N_{ds}$ is divided here to penalize different ranks."}
The performance evaluation is discussed with the plot of BER or RawBER versus signal to noise ratio in terms of $E_b/N_o$ in the dB scale, where each BER point or RawBER point is tested with $100$ subframes.

%%%%%%%%%%%%%%%%%%%%%%%%%%%%%%%%%%%%%%%%%%%%%%%%%%%%%%%%%%%%%%%%%%%%%%%%%%%%
\subsection{Comparison with the State-of-Art Detection Strategies}

To evaluate the performance under the task of symbol detection, we first compare RC-Struct with different approaches in terms of the BER without transmission adaptation. 
% We compare the methods under two conditions: channel without power amplifier (PA) distortion (linear region) and channel with PA distortion (the non-linear region).

% under the linear region in Figure \ref{figs:no_adapt_results} and the non-linear region in Figure \ref{figs:no_adapt_nonlinear_results}.
% We report the RawBER curve of different approaches in Figure \ref{figs:linear_results}.
% We report the RawBER curve of different approaches under the linear region in Figure \ref{figs:linear_results} and the non-linear region in Figure \ref{figs:non_linear_results}.
% While MMNet has achieved great performance, it requires a huge amount of training data and larger computation complexity.

% The results illustrate that the RC-Struct can outperform all the conventional methods with a large margin.
% When compared with RC-Net, RC-Struct still show its advantage in the low $Eb/No$ regime. 
% In Fig.~\ref{figs:channel_linear_rank} and Fig.~\ref{figs:channel_linear_throughput}, we present the statistics of the rank and capacity for the $100$ channel realizations after rank and link adaptation.
% Fig.~\ref{figs:channel_linear_rank} and Fig.~\ref{figs:channel_linear_throughput} illustrate that in the low Eb/No regime, data is transmitted with lower rank and modulation order. \jiarui{expand here}

% \input{figures/figs_channel_linear_throughput}

\begin{figure}
\centering
\includegraphics[width=0.7\linewidth]{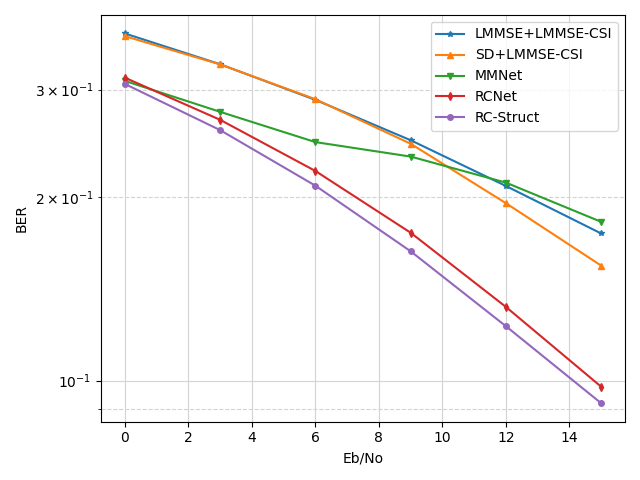}
\caption{Comparison of BER in the linear region.}
\label{figs:no_adapt_results}
\end{figure}

In Fig.~\ref{figs:no_adapt_results}, we show the BER as a function of $E_b/N_o$ for various symbol detection methods. %in the the linear region where no PA distortion exists.
As indicated from the results, both RC-based methods achieve better symbol detection performance compared with conventional model-based methods, which highlights the effectiveness of RC-based schemes in the linear region of the channel.
As the inaccurate LMMSE channel estimation in the low $E_b/N_o$ regime affects the performance of SD, the SD method is shown to have close performance with the LMMSE detection approach.
In addition, RC-Struct is shown to outperform RCNet across all evaluated $E_b/N_o$ values.
This is because the equalization after RC works as the nearest neighbor detector.
RC-Struct, instead, works as a non-linear detector, making it better to classify the data.
We will discuss this with more details in Section~\ref{sec_eval:nn_analysis}.

% While MMNet has shown great performance when there is sufficient amount of training symbols (e.g. 500 pilot symbols with perfect CSIs for training) as proved in~\cite{khani2020adaptive}, it suffers from the online scenario where there are only $4$ pilot symbols and estimated CSI.
While MMNet has shown impressive performance in~\cite{khani2020adaptive}, it is noteworthy that $500$ pilot symbols with perfect CSIs are utilized for the underlying training, which is different from the online setup used in this paper.
In this paper, we conduct the detection subframe by subframe, which is an online over-the-air scenario.
The number of pilot symbols is set as $4$ per subframe, which is far less than the setting in~\cite{khani2020adaptive}. 
Note that the total number of trainable parameters in MMNet is around 40K, while the number of trainable parameters in RC-Struct is around 7K.
As MMNet has more learnable weights, it is shown to be less effective than RC-Struct when learned with only a limited number of training data, as shown in Fig.~\ref{figs:no_adapt_results}.
In addition, the BER of MMNet does not improve much as $E_b/N_o$ increases.
As a large number of weights need to be learned, the MMNet is more likely to suffer from overfitting in the high $E_b/N_o$ regime when the training data is limited.
% Meanwhile, the computation cost for MMNet is much higher than that for RC-Struct.
% RC-Struct trains $7$ RBGs with a single $2$-layer NN, while MMNet requires $N_{\mathrm{sc}}$ $10$-layer NNs for a subframe.
% We show that RC-Struct outperforms MMNet in all values of $E_b/N_o$ in Fig.~\ref{figs:no_adapt_results}.

\subsection{Comparison of Strategies under system non-linearity}

To investigate the performance of the approaches under system non-linearity, we incorporate the following PA model in the evaluation~\cite{rapp1991effects}:
\begin{align}
\phi(x) = \frac{x}{\left[1 + \left(\frac{|x|}{x_{\mathrm{sat}}}\right)^{2\rho}\right]^{0.5\rho}},
\end{align}
where $x$ is the input transmission signal, $\rho$ represents the smoothing parameter, and $x_{\mathrm{sat}}$ measures the saturation level.
The function indicates that when $|x| \ll x_{\mathrm{sat}}$, the $\phi(x)$ is approximately equal to $x$, which forms the linear region without distortion. 
When it comes to the region where $x$ approximates $x_{\mathrm{sat}}$, the function becomes non-linear and the input signal is distorted.
In a nutshell, the distortion occurs when the peak-to-average-power-ratio (PAPR) of the input signal is higher than the input back-off (IBO), where the IBO is the ratio between PA’s saturation power to the input power. 
In this paper, the parameters are set as $x_{\mathrm{sat}} = 1$ and $\rho = 3$.
As PAPR is controlled in the range of $6$ dB to $9$ dB, the non-linear region is chosen by setting IBO smaller than $6.5$ dB.

\begin{figure}
\centering
\includegraphics[width=0.7\linewidth]{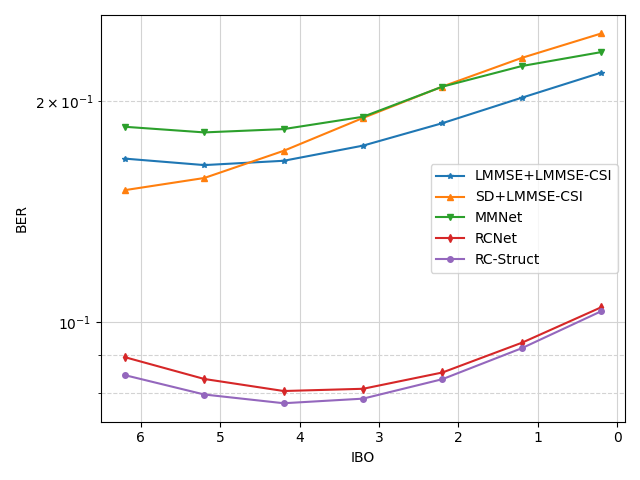}
\caption{Comparison of BER in the non-linear region.}
\label{figs:no_adapt_nonlinear_results}
\end{figure}

In Fig.~\ref{figs:no_adapt_nonlinear_results}, we compare BER for different approaches when PA non-linear distortion occurs.
% versus the input back-off (IBO) plot for different approaches in the non-linear region, where IBO is the ratio between PA’s saturation power to the input power.
Fig.~\ref{figs:no_adapt_nonlinear_results} demonstrates that the RC-based approaches can effectively combat the non-linearity of the PA.
The RC-Struct continues showing advantages over all other methods. 
In the low IBO regime, the performance gap between RC-Struct and RCNet becomes smaller.
This is because the low IBO makes the transmitted signal severely distorted, which affects the LMMSE-based shifting parameter and results in the poor performance of RC-Struct.
The MMNet collapses as it models system without non-linearity and is affected by inaccurate LMMSE-based CSI in non-linear region.

% In Figure \ref{figs:channel_linear_stats} and Figure \ref{figs:channel_nonlinear_stats}, we present the statistics of the rank and capacity for the $100$ channel realizations in the linear region and non-linear region after rank and link adaptation.
% Figure \ref{figs:channel_linear_stats} illustrate that in the low Eb/No region, data is transmitted with lower rank and modulation order.
% Similarly, when IBO is high, the channels tends to have lower rank and capacity, as conveyed in Figure \ref{figs:channel_nonlinear_stats}.

%%%%%%%%%%%%%%%%%%%%%%%%%%%%%%%%%%%%%%%%%%%%%%%%%%%%%%%%%%%%%%%%%%%%%%%%%%%%
\subsection{Effectiveness of Structure-based Neural Network}
\label{sec_eval:nn_analysis}

In this section, we discuss the effectiveness of the classification network in the frequency domain and analyze the performance comparison with more details. 
For ease of discussion, we choose to analyze the classification results with 16-QAM modulation at $15$ dB and $5$ dB $E_b/N_o$ for the $4 \times 4$ MIMO-OFDM system without rank and link adaptation.
Note that RCNet processes the received signals with the RC-based time domain data-stream decoupling and equalization, following with a frequency domain nearest-neighbor classification.
RC-Struct replaces the nearest-neighbor classification with a frequency domain structure-based NN.

\begin{figure}
\centering%
\subfloat[]{\label{a}\includegraphics[width=3cm]{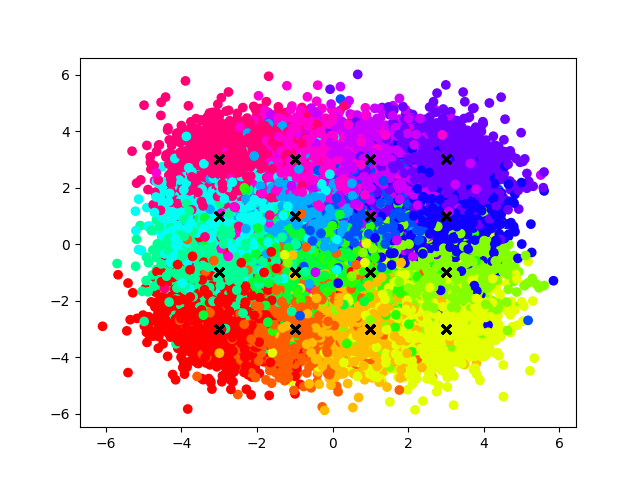}}
\subfloat[]{\label{b}\includegraphics[width=3cm]{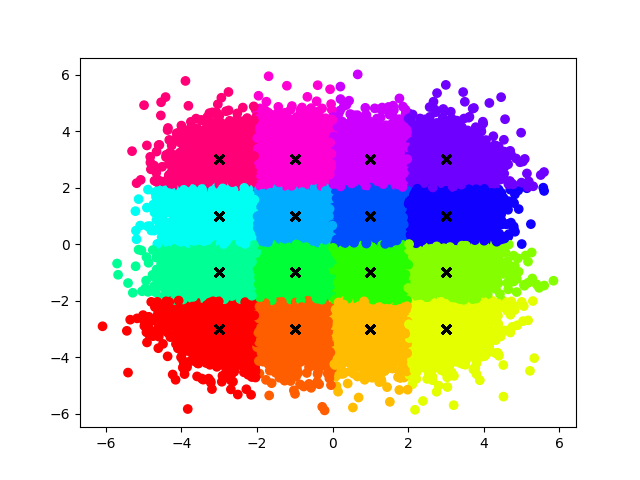}}
\subfloat[]{\label{c}\includegraphics[width=3cm]{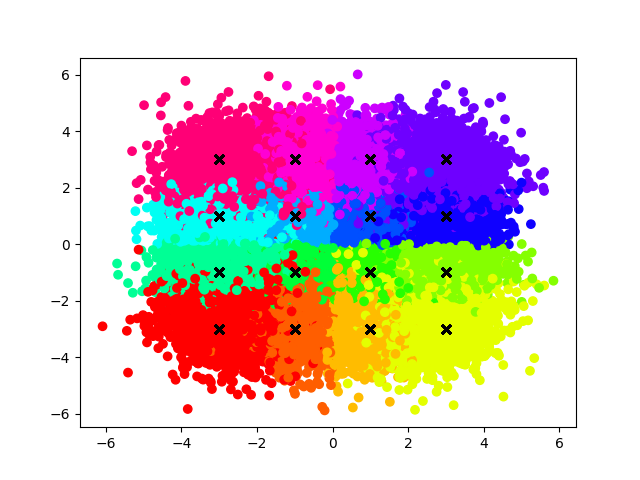}}
\caption{Constellation classification for the RC processed received symbols at $E_b/N_o = 15$ dB with 16-QAM modulation. 
% Each color represents for a constellation class. The cross-marks represent for the exact constellation point positions.
(a) Colored with ground truth labels
(b) Colored with RC equalized labels
(c) Colored with RC-Struct predicted labels
\label{figs:const_compare_15db}
}
\end{figure}

In Fig.~\ref{figs:const_compare_15db} (a), we show the signal constellation of the frequency domain received symbols after RC-based time domain data-stream decoupling and equalization.
%the equalized constellation labels for the RC processed received symbols. 
Each color in the plot represents a particular constellation class ($16$ constellation classes in total) and the cross-marks represent the original position of the transmitted signal constellation point. 
Fig.~\ref{figs:const_compare_15db} (b) shows the frequency domain decision boundary of RCNet after RC-based decoupling and equalization in the time domain and nearest-neighbor classifier in the frequency domain.
% It can be seen that it classifies the received signal constellation to the closest transmitted signal constellation.
The nearest-neighbor method classifies the received signal constellation to the closest transmitted signal constellation.
Comparing Fig.~\ref{figs:const_compare_15db} (a) and (b), we can see that the ground truth labels for the received symbols are not necessarily scattered in the nearest region of the corresponding transmitted constellation point and there are still noise within the received symbols.
By applying the frequency domain NN, RC-Struct introduces non-linear decision boundaries in the classification processes, as shown in Fig.~\ref{figs:const_compare_15db} (c).
Although the decision boundaries are still not perfectly aligned with the ground truth decision boundaries, introducing more non-linearity in the classifier has made the decision boundaries closer to the ground truth ones.

\begin{figure}
\centering%
\subfloat[]{\label{a}\includegraphics[width=3cm]{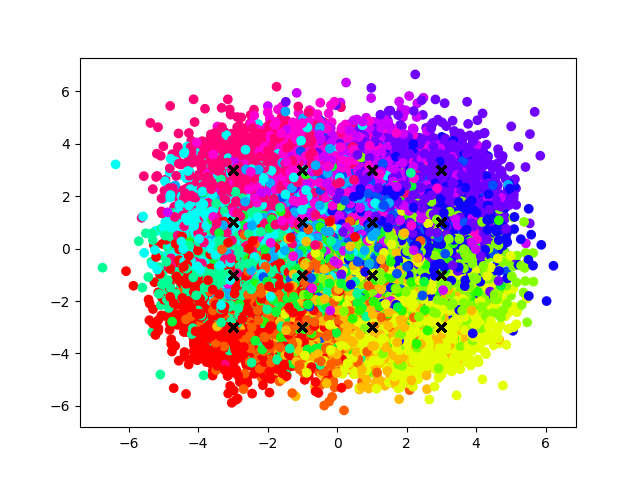}}
\subfloat[]{\label{b}\includegraphics[width=3cm]{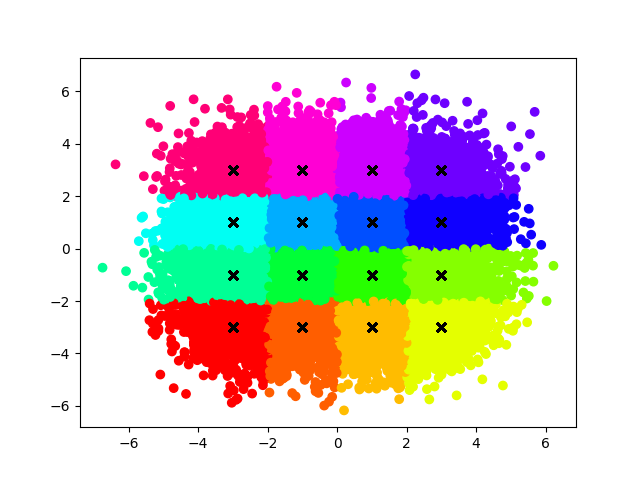}}
\subfloat[]{\label{c}\includegraphics[width=3cm]{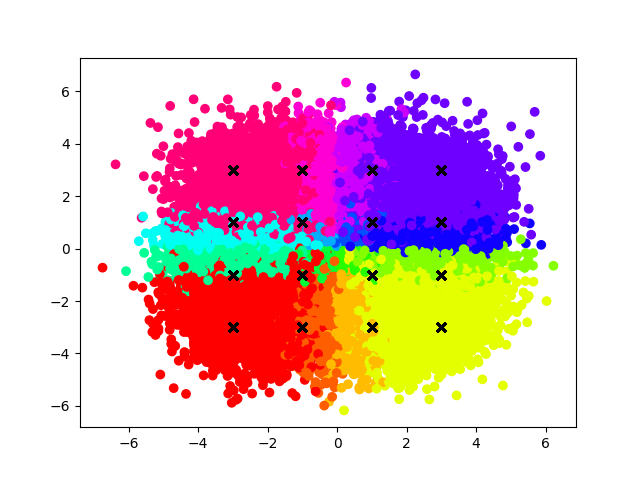}}
\caption{Constellation classification for the RC processed received symbols at $E_b/N_o = 5$ dB with 16-QAM modulation. 
% Each color represents for a constellation class. The cross-marks represent for the exact constellation point positions.
(a) Colored with ground truth labels
(b) Colored with RC equalized labels
(c) Colored with RC-Struct predicted labels
\label{figs:const_compare_5db}
}
\end{figure}

When the $E_b/N_o$ decreases to $5$ dB, the equalized received signals are noisier as shown in Fig.~\ref{figs:const_compare_5db} (a).
It is clear that received symbols cannot be well separated by the linear boundaries, leading to a poor performance of RCNet where the decision boundary between classes is linear.
%learnt boundary line between classes tends to be linear.
% The noisy symbols cannot be well separated by a nearest neighbor classifier, leading to a poor performance of RCNet.
%With the neural network employed in the frequency domain, 
As exhibited in Fig.~\ref{figs:const_compare_5db} (c), RC-Struct learns more noise-tolerable boundary lines to separate data and is shown to be more effective than the linear separation of RCNet in Fig.~\ref{figs:no_adapt_results}.
However, both methods are affected by the noise in the low $E_b/N_o$ regime, resulting in high BER performance.
% As the $E_b/N_o$ further decreases, the data becomes noisier and noisier, making it difficult to be separated with linear boundaries.
% Therefore, the RC-Struct performs better than RCNet especially in the low SNR regime.

% \begin{figure}
% \centering
% \includegraphics[width=0.7\linewidth]{results_figs/cond_num_hist.png}
% \caption{Histogram of channel condition number.}
% \label{figs:channel_condition_hist}
% \end{figure}

% \begin{figure*}
% \includegraphics[width=\ftqa]{results_figs/prev_cond_num_hist_rankNone_ebno5.png}\hfill%
% \includegraphics[width=\ftqa]{results_figs/adapted_cond_num_hist_rankNone_ebno5.png}
% \caption{Histogram of channel condition number.
% \label{figs:channel_condition_hist}
% }
% \end{figure*}

\begin{figure}
\centering%
\subfloat[]{\label{a}\includegraphics[width=4.5cm]{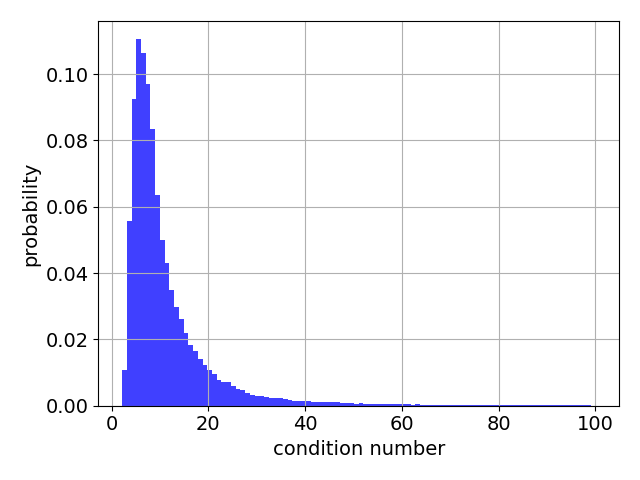}}
\subfloat[]{\label{b}\includegraphics[width=4.5cm]{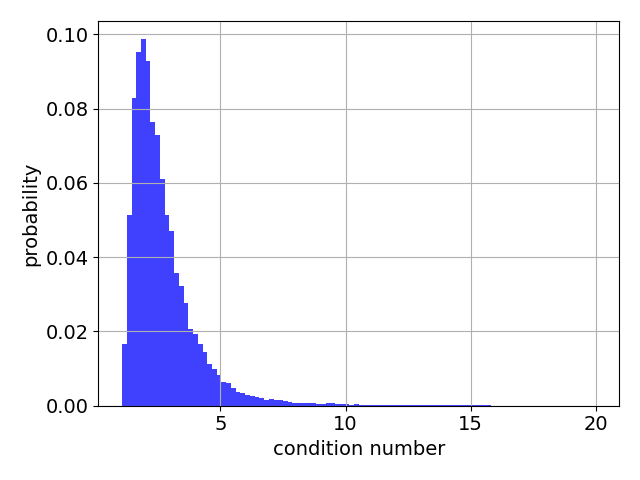}}
\caption{Histogram of channel condition number. 
(a) before rank adaptation.
(b) after rank adaptation
\label{figs:channel_condition_hist}
}
\end{figure}

% \begin{figure}
% \begin{center}
%   %
% \parbox[t]{4cm}{
% \centering%
% \includegraphics[width=6cm]{results_figs/prev_cond_num_hist_rankNone_ebno5.png}\\%
%  \footnotesize (a)}%
% \hspace{2cm}
% %
% \parbox[t]{4cm}{
% \centering%
% \includegraphics[width=6cm]{results_figs/adapted_cond_num_hist_rankNone_ebno5.png}\\%
%  \footnotesize (b)}%

% \end{center}
% \caption{
% Channel condition number histogram. 
% (a) condition number before rank adaptation.
% (b) condition number after rank adaptation
% }
% \label{figs:channel_condition_hist}
% \end{figure}

% \input{figures/figs_constellation_rc_15db}
% \input{figures/figs_constellation_gt_15db}
% \input{figures/figs_constellation_rc_struct_15db}

%%%%%%%%%%%%%%%%%%%%%%%%%%%%%%%%%%%%%%%%%%%%%%%%%%%%%%%%%%%%%%%%%%%%%%%%%%%%
\subsection{Performance comparison with transmission adaptation}

Due to the channel variation and user mobility, the condition number of the underlying MIMO channel in a mobile broadband network such as 4G and 5G/5G-Advanced is also changing rapidly. 
Transmitting with full rank and fixed modulation constellations will result in poor system performance. 
In Fig.~\ref{figs:channel_condition_hist}, we show the channel condition numbers before and after rank adaptation for a duration of $100$ milliseconds of a 3GPP 3D MIMO channel~\cite{study3d3gpp}.
%the condition number for the effective channels
%and after rank adaptation.
Without rank adaptation, the condition number is calculated as the ratio of the largest singular value of the MIMO channel in the frequency domain to its smallest singular value.
With rank adaptation, the condition number is calculated based on the effective MIMO channel after precoding.
%Note that the underlying MIMO channel is generated using the QuaDRiGa model~\cite{jaeckel2014quadriga} for a duration of $300$ milliseconds following the 3GPP 3D MIMO model~\cite{study3d3gpp}.
As shown in Fig.~\ref{figs:channel_condition_hist}, without rank adaptation, most of the MIMO channels have a relatively high condition number.
However, after rank adaptation, the underlying effective MIMO channels have much smaller condition numbers.
% As shown in Fig.~\ref{figs:channel_condition_hist} (a), the 
This is the reason why all detection strategies presented in Fig.~\ref{figs:no_adapt_results} are shown to have higher BERs, when MIMO transmission adaptation is not adopted.
In fact, this is also why link adaptation and rank adaptation are introduced in 4G and 5G/5G-Advanced networks as important features to support MIMO communications~\cite{4GMIMO_OFDM}.

\begin{figure*}
\centering%
\subfloat[]{\label{a}\includegraphics[width=5.5cm]{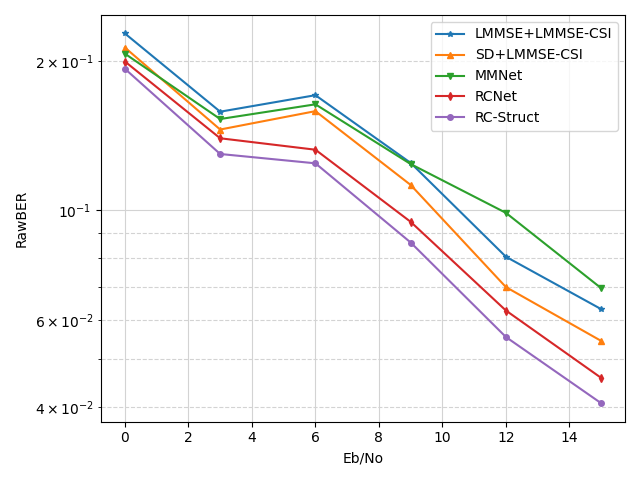}}
% \hspace{1cm}
\subfloat[]{\label{b}\includegraphics[width=5.5cm]{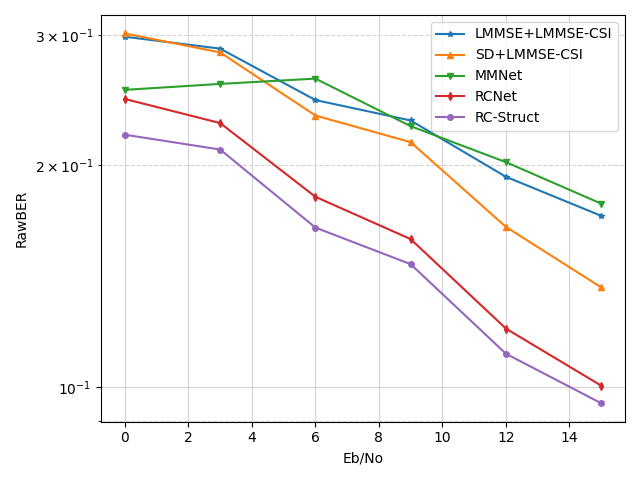}}
% \hspace{1cm}
\subfloat[]{\label{b}\includegraphics[width=5.5cm]{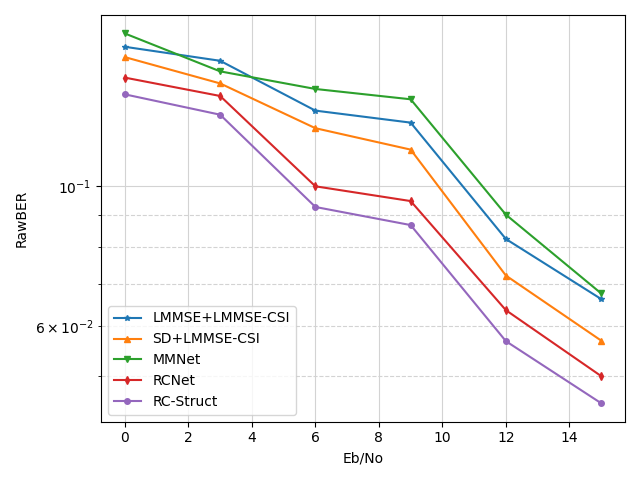}}
\caption{Comparison of RawBER with adaptation.
(a) rank adaptation only.
(b) link adaptation only.
(c) rank and link adaptation.
\label{figs:adaptation_results}
}
\end{figure*}

%To mitigate this issue and validate the advantages of the introduced method under different scenarios, 
In this section, we analyze the performance of various symbol detection methods under link and rank adaptation.
%are utilized in the system.
It is important to note that this is a critical step to evaluate the promise of NN-based symbol detection in realistic and meaningful mobile broadband networks.
To demonstrate the effectiveness of link adaptation and rank adaptation, we provide performance analysis when link adaptation and rank adaptation are utilized separately.
We can show that our RC-based approaches preserve their advantages in all cases as they do not rely on explicit system modeling.
The procedures of link and rank adaptation are detailed in the Appendix.
% As MMNet requires offline weights to perform well, which is not able to be obtained with link and rank adaptation on the fly, we do not compare with MMNet in this section.

In the experiments, we follow current 4G/5G standards to conduct the corresponding link and rank adaptation.
Specifically, the link adaptation will modulate the data streams using either QPSK, $16$-QAM, or $64$-QAM following 4G/5G standards~\cite{std3gpp36211}.
The modulation order is adjusted in a wide-band fashion and the reference SINR is referred to~\cite{link_adapt_table}.
For the rank adaptation, the precoding matrix is obtained by conducting the singular value decomposition (SVD) of the LMMSE estimated channel matrix.
% The rank for transmission is adjusted across all the subcarriers
The rank of the MIMO transmission is adjusted across all the subcarriers complying with the wideband-based rank indicator (RI) feedback~\cite{std3gpp36213}, and the precoding matrix is applied in a group of subcarriers following the procedure of the subband-based precoding matrix indicator (PMI) feedback~\cite{std3gpp36213}.
The sub-band size is set as $84$ subcarriers, which is $7$ RBGs.
The detailed procedure is described in the Appendix.

% \input{figures/figs_adaptation}
% \input{figures/figs_rank_adapt_only}

% when most of the channels have a relative high condition number, as shown in Fig.~\ref{figs:channel_condition_hist} (a). 

\begin{figure}
\centering
\includegraphics[width=4cm]{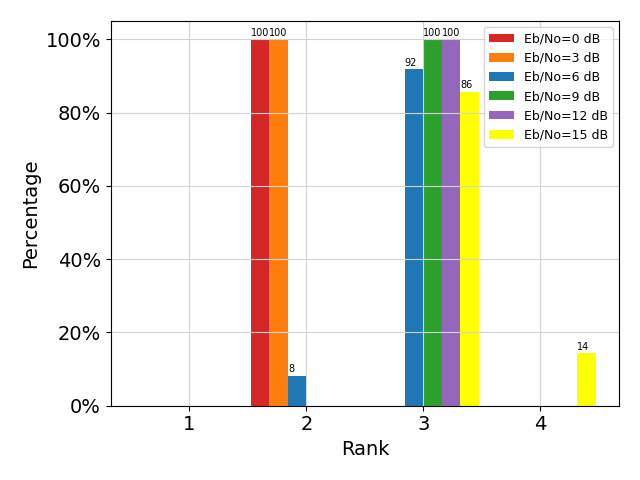}%
\includegraphics[width=4cm]{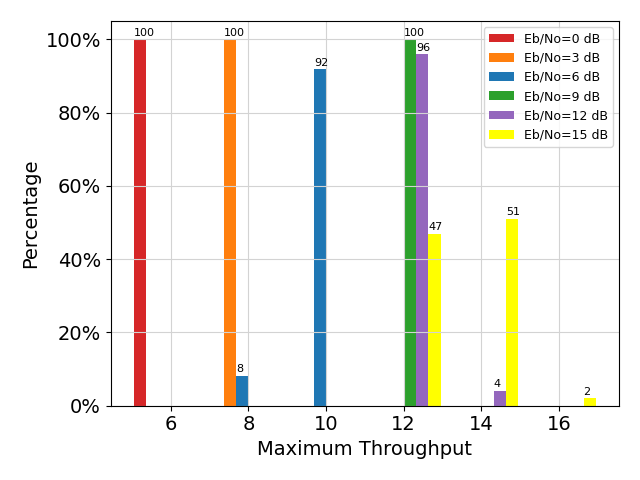}
% \hfill%
% \includegraphics[width=\ftqa]{results_figs/channel_qam_linear.png}
\caption{Percentage of adapted rank and capacity for all the tested channels.
\label{figs:channel_linear_stats}
}
\end{figure}

% When rank adaptation is conducted, the underlying effective MIMO channels have much smaller condition numbers, as shown in Fig.~\ref{figs:channel_condition_hist} (b).
% The effective MIMO channel here refers to the precoded MIMO channels.
When rank adaptation is conducted, the RawBER for all the methods decreases compared with the case without utilizing adaptation, as shown in Fig.~\ref{figs:adaptation_results} (a).
Note that in this case, RawBER is equivalent to BER since link adaptation is not utilized.
RC-Struct continues demonstrating its advantage over all other methods.
These results clearly underline the effectiveness of RC-Struct in both good and bad channel conditions.
For LMMSE, SD, and MMNet, the RawBER at $3$ dB $E_b/N_o$ is lower than $6$ dB $E_b/N_o$.
As illustrated in Fig.~\ref{figs:channel_linear_stats}, the channels at $0$ dB and $3$ dB $E_b/N_o$ are all adapted to rank $2$, while those at $6$ dB have a high percentage to be adapted to rank $3$.
Therefore, the RawBER curve for these three methods from $0$ dB to $3$ dB $E_b/N_o$ is decreasing since the adapted rank is the same, while the curve increases from $3$ dB to $6$ dB $E_b/N_o$ as the channels have a higher percentage to be adapted to higher ranks.
% This is because lower rank is adapted at $E_b/N_o$ $3$ dB as shown in Fig.~\ref{figs:channel_linear_stats}, making it easier for the LMMSE, SD, and MMNet methods to detect.
% On the other hand, the performance margin between RC-Struct and RCNet becomes smaller in the low $E_b/N_o$ regime.
% This is because the frequency domain received symbols after RC-based time decoupling and equalization tends to be easier to be separated with linear boundaries when MIMO channels have small condition numbers.
% The RawBER curve for MMNet is not monotonically decreasing in this scenario, as it is highly affected by the adapted rank.
% As illustrated in Fig.~\ref{figs:channel_linear_stats}, the channels at $0$ dB and $5$ dB $E_b/N_o$ are all adapted to rank $2$, while those at $10$ dB and $15$ dB have higher percentage to be adapted to higher rank.
% Therefore, the RawBER curve for MMNet from $0$ dB to $5$ dB $E_b/N_o$ is decreasing since the adapted rank is the same, while the curve increases from $5$ dB to $15$ dB $E_b/N_o$ as the channels have higher percentage to be adapted to higher ranks.
% On the other hand, the MMNet model collapses when rank adaptation is conducted, as the LMMSE estimated effective channel is not accurate. 
When link adaptation is applied, RC-Struct preserves its advantage over the other methods, as shown in Fig.~\ref{figs:adaptation_results} (b).
% Employing link adaptation shows smaller performance improvement than rank adaptation, as link adaptation does not change the channel conditions.

% \input{figures/figs_link_adapt_only}

When employing both link and rank adaptation in the system, the performance of all the methods has been improved, as shown in Fig.~\ref{figs:adaptation_results} (c).
The performance improvement is because the transmission rank and modulation order are dynamically adapted, as presented in Fig.~\ref{figs:channel_linear_stats}.
The dynamically adapted rank and modulation order, on the other hand, leads to the zigzag pattern of the performance curve, as lower transmission rank and modulation order may be adopted in lower $E_b/N_o$ regimes.
% The non-monotonic curve for MMNet is affected by the rank adaptation as illustrated above.
% In Fig.~\ref{figs:channel_linear_rank} and Fig.~\ref{figs:channel_linear_throughput}, we provide the percentage of adapted rank and the percentage of adapted maximum throughput under different $E_b/N_o$ respectively.
% In Fig.~\ref{figs:channel_linear_stats}, we provide the percentage of adapted rank and the percentage of adapted maximum throughput under different $E_b/N_o$ respectively.
% As presented in Fig.~\ref{figs:channel_linear_stats}, in the low SNR regime, data is transmitted with lower rank and modulation order, making the frequency domain received signal after RC-based time decoupling and equalization easier to be separated with linear boundaries and resulting in the close performance between RCNet and RC-Struct in the low SNR regime.
RC-Struct still achieves the best performance among all the compared methods and demonstrates its ability to be applied with dynamic transmission modes.
Furthermore, we conducted an experiment when LDPC coding is adopted with link and rank adaptation at $15$ dB $Eb/N_o$.
In 3GPP 5G NR~\cite{std3gpp38214}, the code rate ranges from $0.0762$ to $0.9258$.
In our experiment, the code rate for LDPC coding is set as $0.3125$.
Tab.~\ref{tab:ber_ldpc_coding} shows that the BER is $1.84\%$ and BLER is $6.45\%$ after LDPC.
This result demonstrates that RC-Struct can meet the target BLER of $10\%$ as specified in 3GPP 5G NR~\cite{std3gpp38214}.

% \newlength\ftqa
% \setlength\ftqa{7.0cm}
% \newlength\ftqb
% \setlength\ftqb{0.1mm}
% \newlength\ftqc
% \setlength\ftqc{0.8mm}

Here are the summaries and takeaways for the experiments. 
\begin{itemize}
\item When no adaptation is applied, RC-Struct outperforms all existing methods.
    % Specifically, when compared with RCNet, it shows 
It is shown to have lower BER than RCNet, due to its ability to handle non-linear boundaries in the frequency domain.
\item For systems with unknown PA non-linearity, RC-Struct and RCNet shows relatively large performance gains over other methods.
In the low IBO regime, the performance gap between RC-Struct and RCNet becomes smaller since the severely distorted signal negatively affects the estimation of the shifting parameter.
\item When rank adaptation and link adaptation are conducted, RC-Struct and RCNet can effectively conduct online learning with extremely limited training symbols to adjust detection strategies on a subframe-by-subframe basis. 
Meanwhile, RC-Struct continues showing advantages over other methods including RCNet.
When LDPC coding is adopted, RC-Struct has a BLER of $6.45\%$, which is lower than the target BLER specified in~\cite{std3gpp38214}.
% When both rank and link adaptation are conducted, RC-Struct share similar RawBER performance as RCNet and outperforms all other detection strategies.
\end{itemize}

\begin{table}[t]
\centering
\caption{Performance when adopted LDPC channel coding at $15$ dB $E_b/N_o$.
}
\label{tab:ber_ldpc_coding}
{
\begin{tabular}{l|cc}
\toprule
%  & BER & Coded BER & BLER \\
% \midrule
% RC-Struct & $5.65\%$ & $1.84\%$ & $6.45\%$ \\
 & Coded BER & BLER \\
\midrule
RC-Struct & $1.84\%$ & $6.45\%$ \\
\bottomrule
\end{tabular}
}
\end{table}

\section{Conclusion}
\label{conclusion}
In this paper, we introduced an NN-based symbol detector for MIMO-OFDM systems, which incorporates the structural knowledge of the systems, including the temporal dynamics within the data, the time-frequency structure of the OFDM waveform, and the repetitive structure of the modulation constellation.
The temporal information is captured by the RC network in the time domain, while the constellation symmetry is leveraged by the classifier in the frequency domain.
% The channel state information (CSI) is learned dynamically by the network.
The network architecture allows the network to be learned with significantly reduced training overhead and can be applied with rank adaptation and link adaptation of the underlying MIMO transmissions.
Experiments indicate the effectiveness of the introduced RC-Struct network and demonstrate the advantages of incorporating the structure information into the network under MIMO channels with time-dynamic transmission modes.
Due to the ability to be applied in practical MIMO operations, RC-Struct provides a promising symbol detection approach for the MIMO-OFDM system in the 5G/5G-Advanced and Beyond networks.

% \input{figures/figs_channel_linear_throughput}

%%%%%%%%%%%%%%%%%%%%%%%%%%%%%%%%%%%%%%%%%%%%%%%%%%%%%%%%%%%%%%%%%%%%%%%%%%%%
%%%%%%%%%%%%%%%%%%%%%%%%%%%Appendix%%%%%%%%%%%%%%%%%%%%%%%%%%%%%%%%%%%%%%%%
%%%%%%%%%%%%%%%%%%%%%%%%%%%%%%%%%%%%%%%%%%%%%%%%%%%%%%%%%%%%%%%%%%%%%%%%%%%%

% if have a single appendix:
%\appendix[Proof of the Zonklar Equations]
% or
%\appendix  % for no appendix heading
% do not use \section anymore after \appendix, only \section*
% is possibly needed

% use appendices with more than one appendix
% then use \section to start each appendix
% you must declare a \section before using any
% \subsection or using \label (\appendices by itself
% starts a section numbered zero.)
%
\appendices
\section{Rank adaptation and link adaptation}
\label{appendix:rank_link_adaptation}
\subsection*{Rank Adaptation}
% \label{rank_adaptaion}
% \jiarui{I feel we need to mention something about 5G systems.}
A key premise of 5G/5G-Advanced system is its ability to dynamically switch between different ranks to efficiently exploit the bandwidth for a given channel condition and improve the performance of the system, since there is no single mode that works best in all channel conditions\cite{sheikh2011performance}.
The number of transmitted data streams is referred to as the transmission rank.
We focus on the capacity based rank adaptation to adjust the rank of the wireless channel.
% Rank of the MIMO channel is defined as the number of independent data streams that can be reliably transmitted through the channel.

% To fully take advantage of the space dimension flexibility, the precoding matrix is applied to achieve different streams transmission.
The precoding process for $L$ data streams transmission can be represented as $\boldsymbol{X}_n(k) = \boldsymbol{Q}_n(k)\boldsymbol{S}_n(k)$, where $\boldsymbol{Q}_n(k) \in {\mathbb C}^{N_t \times L}$ is the unitary precoding matrix and $\boldsymbol{S}_n(k) \in {\mathbb C}^{L \times 1}$ is the effective transmitted symbols. 
Suppose the precoding matrix is $\boldsymbol{Q}_n(k) = \boldsymbol{V}_n^L(k)$, where $\boldsymbol{V}_n^L(k)$ is first $L$ columns from the unitary matrix $\boldsymbol{V}_n(k)$ in the SVD $\boldsymbol{H}_n(k) = \boldsymbol{U}_n(k)\boldsymbol{\Lambda}_n(k)\boldsymbol{V}_n(k)^H$. 
The throughput of $L$ data streams transmission can be written as \cite{zhang2011precoding}
\begin{align}
C_L = \sum_{l=1}^L \log_2 \left( 1 + \frac{P_t}{L\sigma^2}\lambda_l^2 \right),
\end{align}
where $P_t$ is the total transmit power and $\lambda_l$ is the $l$th singular value in the SVD of $\boldsymbol{H}_n(k)$ with the order of $\lambda_{max} = \lambda_1 \geq \dots \geq \lambda_{min} \geq 0$. 
The rank is adapted to take the maximum throughput among all the possible $L$ values ($L \leq N_t$). 

%%%%%%%%%%%%%%%%%%%%%%%%%%%%%%%%%%%%%%%%%%%%%%%%%%%%%%%%%%%%%%%%%%%%%%%%%%%%%%%%%%%%%%%%%%%%%%%%%%%%%%%%%%%%%%%%%%%%%%%%%%%%%%

\subsection*{Link Adaptation}
% \label{link_adaptaion}
% \jiarui{may need to check the accuracy.}
In 5G wireless communication network, multiple modulation and coding schemes (MCS) are enabled to transmit with higher data rates and reliability. 
The transmitter is expected to transmit data with a proper MCS according to the channel conditions.
Link adaptation techniques select MCS for wireless transmission based on the channel quality indicator (CQI) feedback.
In a simple case, the base station (BS) transmits pilot signals to user equipment (UE) at given OFDM subcarrier positions.
The UE measures the SINR at each OFDM subcarrier and calculates an effective SINR using techniques such as effective exponential SNR mapping (EESM)\cite{sandanalakshmi2007effective}.

The EESM approach has been widely applied to the OFDM link layers. 
It maps individual subcarrier SINRs to an effective SINR with the following equation:
\begin{align}
SINR_{eff} = -\beta ln\left[ \frac{1}{S} \sum_{n=1}^{S} e^{-\frac{SINR_n}{\beta}}\right],
\end{align}
where $SINR_n$ is the $n$th subcarrier SINR and $S$ represents the size of the subband. 
The parameter $\beta$ is empirically obtained and is calibrated to fit the model for different MCS level.
The effective SINR is compared with the reference SINR value and mapped to a CQI value, which indicates the highest modulation order and code rate for keeping a Packet Error Rate (PER) below $10\%$~\cite{link_adapt_table}.
Once the CQIs are collected by the base station, it allocates resources for each user.

\ifCLASSOPTIONcaptionsoff
  \newpage
\fi

% trigger a \newpage just before the given reference
% number - used to balance the columns on the last page
% adjust value as needed - may need to be readjusted if
% the document is modified later
%\IEEEtriggeratref{8}
% The "triggered" command can be changed if desired:
%\IEEEtriggercmd{\enlargethispage{-5in}}

% references section

% can use a bibliography generated by BibTeX as a .bbl file
% BibTeX documentation can be easily obtained at:
% http://mirror.ctan.org/biblio/bibtex/contrib/doc/
% The IEEEtran BibTeX style support page is at:
% http://www.michaelshell.org/tex/ieeetran/bibtex/
\bibliographystyle{IEEEtran}
% argument is your BibTeX string definitions and bibliography database(s)

\bibliography{IEEEabrv,ref.bib}
%
% <OR> manually copy in the resultant .bbl file
% set second argument of \begin to the number of references
% (used to reserve space for the reference number labels box)

% biography section
% 
% If you have an EPS/PDF photo (graphicx package needed) extra braces are
% needed around the contents of the optional argument to biography to prevent
% the LaTeX parser from getting confused when it sees the complicated
% \includegraphics command within an optional argument. (You could create
% your own custom macro containing the \includegraphics command to make things
% simpler here.)
%\begin{IEEEbiography}[{\includegraphics[width=1in,height=1.25in,clip,keepaspectratio]{mshell}}]{Michael Shell}
% or if you just want to reserve a space for a photo:

% \begin{IEEEbiography}{Michael Shell}
% Biography text here.
% \end{IEEEbiography}

% % if you will not have a photo at all:
% \begin{IEEEbiographynophoto}{John Doe}
% Biography text here.
% \end{IEEEbiographynophoto}

% insert where needed to balance the two columns on the last page with
% biographies
%\newpage

% You can push biographies down or up by placing
% a \vfill before or after them. The appropriate
% use of \vfill depends on what kind of text is
% on the last page and whether or not the columns
% are being equalized.

%\vfill

% Can be used to pull up biographies so that the bottom of the last one
% is flush with the other column.
%\enlargethispage{-5in}

% that's all folks
\end{document}